\documentclass{aa} 
\usepackage{graphicx}
\usepackage{txfonts}
\usepackage{epsfig}
\usepackage{subfig}
\usepackage{natbib}
\bibpunct{(}{)}{;}{a}{}{,}
\usepackage{color}

\begin{document}
   \title{A Stellar Population Synthesis Model for the Study of Ultraviolet Star Counts of the Galaxy}
  \author{
Ananta~C.~Pradhan\inst{\ref{tifr}}, 
D.~K.~Ojha\inst{\ref{tifr}},
A.~C.~Robin\inst{\ref{besn}},
S.~K.~Ghosh\inst{\ref{tifr},\ref{ncra}},
John~J.~ Vickers\footnote{Member of the International Max-Planck Research School for Astronomy \& Cosmic Physics at the University of Heidelberg}\inst{\ref{ari}}
}

\institute{Tata Institute of Fundamental Research, Homi Bhabha Road,
Mumbai 400005, India\\ \email{acp.phy@gmail.com}
\label{tifr} 
\and 
Institut Utinam, CNRS UMR 6213, OSU THETA, Universit\'{e} de Franche-Comt\'{e}, 41bis avenue de l'Observatoire 25000 Besan\c{c}on, France\label{besn}
\and
 National Centre for Radio Astrophysics, Tata Institute of Fundamental Research, Pune 411007, India\label{ncra}
\and
Astronomisches Rechen-Institut, Zentrum f\"{u}r Astronomie der Universit\"{a}t Heidelberg, M\"{o}nchhofstr, 12-14, 69120 Heidelberg, Germany\label{ari}
}

\date{ }
 
\abstract
   {Galaxy Evolution Explorer (GALEX), the first all sky imaging ultraviolet (UV) satellite, has imaged a large part of the sky providing an excellent opportunity for studying UV star counts. Combining photometry from the different wavelengths in the infrared (from Wide-field Infrared Survey (WISE) and Two Micron All Sky Survey (2MASS)) to UV allows us to extract a real star catalog from the GALEX source catalog.}
{The aim of our study is to investigate in detail the observed UV star counts obtained by GALEX vis-a-vis the model simulated catalogs produced by the Besan\c{c}on model of stellar population synthesis in various Galactic directions, and to explore the potential for studying the structure of our Galaxy from images in multiple near-UV (NUV) and far-UV (FUV) filters of the forthcoming Ultraviolet Imaging Telescope (UVIT) to be flown onboard ASTROSAT.}
{We have upgraded the Besan\c{c}on model of stellar population synthesis to include the UV bands of GALEX and UVIT. Depending on the availability of contiguous GALEX, Sloan Digital Sky Survey (SDSS), WISE and 2MASS overlapping regions, we have chosen a set of nineteen GALEX fields which spread over a range of Galactic directions. We selected a sample of objects from the GALEX database using the \emph{CASjobs} interface and then cross-matched them with the WISE+2MASS and SDSS catalogs. UV stars in the GALEX catalog are identified by choosing a suitable infrared (IR) colour, $J - W1$ (W1 is a WISE band at 3.4 $\mu$m), which corresponds to a temperature range from 1650 K to 65000 K. The IR colour cut method, which is used for the first time for separation of stars, is discussed in comparison with the GALEX+SDSS star counts method.}
{We present the results of the UV star counts analysis carried out using the data from GALEX. We find that the Besan\c{c}on model simulations represent the observed star counts of both the GALEX All-sky Imaging Survey (AIS) and Medium Imaging Survey (MIS) well within the error bars in various Galactic directions. Based on the model analysis, we separated out white dwarfs (WDs) of the disc and blue horizontal branch stars (BHBs) of the halo from the observed sample by selecting a suitable $FUV - NUV$ colour.}
{The Besan\c{c}on model is now ready for further comparisons in the UV domain and will be used for prospective studies for the UVIT instrument to be flown onboard ASTROSAT.}
   
 \keywords{Stars: general - Ultraviolet: stars - Galaxy: disc - Galaxy: stellar content - Galaxy: halo}             
\titlerunning{A Model for the UV Star Counts of the Galaxy}
 \authorrunning{Pradhan et al.}
 \maketitle 

\section{Introduction}

The Milky Way is the best studied Galaxy in the universe; its structure and evolution have been studied by a variety of techniques. In the early 20$^{th}$ century, \citet{Kapteyn22} first studied the geometrical structure of the Galaxy using the star counts method whereby he counted stars on the photographic plates in selected areas of the sky. Since then the star counts method has been used as one of the preferred methods to constrain the structural parameters of the Galaxy effectively. Several reviews \citep{Bahcall86,Freeman87,Gilmore88,Majewski93,Helmi08,Ivezic12} have discussed the connection of star counts to the Galactic structure. The advent of instruments with better resolution and greater sensitivity have enabled us to obtain photometric observations covering large parts of the sky in several wavelength bands. The population synthesis models of the Milky Way are well supported by these observations in predicting the different structural parameters of the Galaxy, such as stellar densities, scale length, scale height, etc. Among the models built to understand the Galactic structure by star counting method, one can cite: \citet{Bahcall80}, \citet{Gilmore83}, \citet{Robin86}, \citet{Robin03}, \citet{Girardi05} and \citet{Juric08}. However, the above Galaxy models are predominantly based on the visible and IR photometric surveys. Very few attempts \citep{Brosch91,Cohen94} had been made to study the star counts in UV prior to GALEX due to a lack of availability of UV photometric surveys. The advent of GALEX, which provided a wide sky coverage in UV, now allows new analysis of the UV sky \citep[][ among others]{Xu05,Bianchi11a,Bianchi11b,Bianchi13}. An attempt has also been made to predict the star counts in the X-ray band \citep{Guillout96} by extending the Besan\c{c}on model of stellar population synthesis \citep{Robin86} to the ROSAT PSPC energy bands.

Indeed, the UV surveys, among others, could help in tracing the spiral structures which mainly contain very young stars. The UV surveys also help in constraining the shape of the initial mass function (IMF) towards the high-mass star end as well as elucidating the recent star formation history. Moreover, they also trace very blue populations such as WDs and BHBs deep in the halo population, which in turn trace the streams and relics of ancient accretion in the Milky Way halo. GALEX has covered a large part of the sky which provides an opportunity to explore and characterize these hot sources in the FUV (1344 - 1786 \AA, $\lambda_\mathrm{eff}$ = 1538.6 \AA) and NUV (1771 - 2831 \AA, $\lambda_\mathrm{eff}$ = 2315.7 \AA) wavebands with better resolution and greater sensitivity than the previous surveys. A vivid description of the source selection, FUV and NUV magnitude error cuts and the statistical analysis of the GALEX catalog is provided by \citet{Bianchi07}, \citet{Bianchi09} and \citet{Bianchi11a, Bianchi13}. Detection of WDs and BHBs is one of the main achievements of GALEX as these sources are elusive in the other wavelength bands of the electromagnetic spectrum due to their high temperature. WDs and BHBs are integral to the study of stellar evolution and structure of the Milky Way as they belong to different stellar populations of the Galaxy.

We have upgraded the Besan\c{c}on model of stellar population synthesis to include the UV bands of GALEX and the upcoming UVIT\footnote{http://www.iiap.res.in/Uvit} (which will be flown onboard ASTROSAT) to predict star counts in different parts of the sky \citep{Todmal10}. UVIT will image the sky in the FUV (1300 - 1800 \AA) and NUV (2000 - 3000 \AA) channels, each having five filters, at a high resolution of \(1.8''\) \citep{Postma11,Kumar12a,Kumar12b}. Better positional accuracy of UVIT as compared to GALEX will enable more reliable cross correlation with other catalogs which will be of great utility in inferring the Galactic structure using the star counts technique. The transmission curves (effective area versus wavelength) for the FUV and NUV bands of GALEX together with each of the five FUV (left panel) and NUV (right panel) filters of the upcoming UVIT/ASTROSAT are shown in Figure 1. We have included the effective area curves of both the GALEX and all the UVIT/ASTROSAT bands in the model to simulate the UV star counts in these bands. Apart from the GALEX bands, we will discuss the model simulated star counts of the BaF2 (FUV: 1370 - 1750 \AA, $\lambda_\mathrm{eff}$ = 1504 \AA) and NUVB4 (NUV: 2505 - 2780 \AA, $\lambda_\mathrm{eff}$ = 2612 \AA) bands of UVIT/ASTROSAT. The expected sensitivity limits (5$\sigma$) in AB magnitude system in the UVIT BaF2 (FUV) and NUVB4 (NUV) wavebands, for an exposure time of 200 seconds, are 20.0 and 21.2 magnitudes, respectively (ASTROSAT Handbook 2013; private communication). 

It is worth mentioning here that throughout the paper we have used AB system for the GALEX, UVIT and SDSS data sets, whereas the 2MASS and WISE data sets are in the Johnson system (see Section 2).

We give details of the observations and selection of UV stars in Section 2. We describe about the Besan\c{c}on model Galaxy model in Section 3 and discuss the comparison of the GALEX+WISE+2MASS and GALEX+SDSS star counts in Section 4. We present the comparison of the model with the observations in Section 5, and discuss the distribution of the model star counts in Section 6. We mention the identification of WDs and BHBs using $FUV - NUV$ colour in Section 7. Finally, we summarize our conclusions in Section 8.

\section{Observations and cross-correlation of GALEX sources}
\subsection{GALEX data}
GALEX was an orbiting space telescope launched in April, 2003, which was terminated in mid-February, 2012. The satellite and on-orbit performance are described in \citet{Martin05} and \citet{Morrissey05,Morrissey07}. It observed the sky in two UV bands, FUV and NUV, simultaneously, with a spatial resolution of 4.2\arcsec\ and 5.3\arcsec, respectively. The field of view is 1.25\degr\ in diameter and the images are sampled with 1.5\arcsec\ pixels. The typical AB magnitude limits (5$\sigma$ depth) met by AIS for an exposure time of 100 seconds and MIS for an exposure time of 1500 seconds are 19.9/20.8 (FUV/NUV) and 22.6/22.7 (FUV/NUV), respectively \citep{Morrissey07}. The AIS has the largest sky coverage when compared to the other GALEX surveys. GALEX has observed a large part of the sky ($\sim$75\%), excepting the Galactic plane and some regions of the Magellanic Clouds due to safety concerns of the detectors. In this paper, we have used the GALEX GR6 data which is available in Multi-mission Archive at Space Telescope Science Institute (MAST\footnote{http://galex.stsci.edu/GR6}). 

\subsection{Selection of GALEX fields}

We have selected nineteen GALEX fields for which both the detectors of GALEX were turned on. We retained sources which had a reliable NUV detection, however, FUV detections are available for $\sim$3.5\% and $\sim$6.8\% of the NUV detections in the selected AIS and MIS fields, respectively. The rest of the NUV sources do not have a FUV detection because their FUV fluxes are too low to be detected. We include only regions within a radius of 0.5\degr\ from the center of the tiles to eliminate edge artifacts and bad sources along the edge as well as to avoid overlapping areas and duplication of the sources. The coverage areas of the observed fields are calculated by summing up the areas of all the tiles in a field.

The fields are selected in the footprints of the GALEX, SDSS, WISE and 2MASS surveys. The various fields are as follows:
\begin{itemize}
\item Four GALEX tiles were chosen at the southern Galactic latitudes: two each in AIS and MIS.
\item Eight fields with large area coverage of the sky were chosen in several northern Galactic directions. The fields include: two regions towards the Galactic center (GC) (one each in AIS and MIS), two regions towards the Galactic anticenter (GAC) (one each in AIS and MIS), and one region each in AIS towards the Galactic antirotation (GAR), Galactic low latitude (GLL), Galactic high latitude (GHL), and Galactic pole (GP) directions.
\item Seven fields in AIS were chosen at 10\degr\ intervals of $b$ around $l \sim$ 50\degr\ in order to study the latitude variation of UV star counts.
\end{itemize}

 The center coordinates, survey types, area coverages, location in the Galaxy, number of GALEX tiles and the range of exposure times of NUV and FUV observations of each of the fields are given in Table 1.

\subsection{WISE+2MASS data}

The AIS and MIS of GALEX overlap with the 2MASS and WISE footprints. 2MASS \citep{Skrutskie06} has observed the entire sky in the J (1.24 $\mu$m), H (1.66 $\mu$m) and K$_{s}$ (2.16 $\mu$m) near-IR (NIR) bands with angular resolutions of \(2.9''\), \(2.8''\), and \(2.9''\) respectively; while WISE \citep{Wright10} has mapped the sky in the W1 (3.4 $\mu$m), W2 (4.6 $\mu$m), W3 (12 $\mu$m), and W4 (22 $\mu$m) mid-IR bands, with angular resolutions of \(6.1''\), \(6.4''\), \(6.5''\), and \(12.0''\) respectively. The 5$\sigma$ point source sensitivities of the four WISE bands are better than 0.08, 0.11, 1 and 6 mJy (equivalent to 16.6, 15.6, 11.3, and 8.0 Vega magnitude) in unconfused regions on the ecliptic \citep{Wright10}. The existing WISE+2MASS cross-matched catalog available at Infrared Science Archive (IRSA\footnote{http://irsa.ipac.caltech.edu/Missions/wise.html}) has been used for convenience. This catalog has been produced using a \(3.0''\) matching radius, which was found to be adequate considering the positional accuracy and resolution.

We have made use of the Virtual Astronomical Observatory (VAO\footnote{http://vao-web.ipac.caltech.edu/applications/VAOSCC}) for cross-matching GALEX sources with WISE+2MASS sources. GALEX sources were uploaded into the VAO, seeking their WISE and 2MASS counterparts using a match radius of \(3.0''\). We found most of the real matched sources within this radius, with a very small fraction ($<$ 1\%) having multiple matches which were removed from the final catalog. We also estimated the possible contamination by spurious matches (random coincidences) for the matched sources following the method of \citet{Bianchi11a}. For this purpose we used a match radius of \(6.0''\), which is equivalent to the resolution of WISE, to find the GALEX counterparts of WISE sources. The spurious matches were found to be $\sim$10\% of the total matched sources, and 75\% of these spurious matched sources lie beyond a distance of \(3.0''\).

\subsection{SDSS data}
So far, SDSS has mapped over 35\% of the full sky in five optical photometric bands ($u, g, r, i, z$) covering the wavelength range from 3000 to 11000 \AA\ \citep{Aihara11}. GALEX GR6 has been cross-matched against SDSS DR7 and the provided cross-matched table is \emph{xSDSSDR7}. Several works \citep{Seibert05,Budavari09,Bianchi07,Bianchi11a} have explained the cross-matching of the GALEX catalog with SDSS, astrophysical source classifications and related statistical analyses. We uploaded the GalexIDs of the objects into the GALEX \emph {CASjobs}\footnote{http://galex.stsci.edu/casjobs} SQL (Structured Query Language) interface to determine their SDSS counterparts in a search radius of 3.0\arcsec. We have eliminated the multiple matches ($<$ 1\%) from the GALEX+SDSS final catalog. The estimated spurious matches in case of GALEX+SDSS are found to be $\sim$7\% within 3.0\arcsec\ radius. The SDSS star/galaxy classifications have been adopted while performing the match in order to separate out point sources from the source list.

SDSS classified point sources (GALEX+SDSS) include both stars and quasi stellar objects (QSOs), out of which we selected QSOs using the SDSS colour cuts from \citet{Richards02} and removed them from the GALEX+SDSS point sources and termed the clean sample as `GALEX+SDSS stars'.

\subsection{Selection of stars from the GALEX catalog by IR colour cut method}

Figure 2 shows $J - W1$ versus NUV colour-magnitude diagram (CMD) of all GALEX sources that are cross-matched with WISE+2MASS sources for the regions in the directions of the GC and the GAC, each covering 69.9 deg$^{2}$ of the sky. QSO candidates are selected using the SDSS colour cuts from \citet{Richards02} and are represented by blue crossed symbols in the plot. We clearly see two groups of sources in the figure well separated by $J - W1$ colour. The stars verified by their SDSS classification as point sources in a cross-matched sample are identified to be bluer than $J - W1 <$ 1.2 and the extra-galactic objects (e.g. galaxies, QSOs, etc.) are redder, with $J - W1 >$ 1.2. Since the contamination by SDSS-identified QSOs is estimated to be negligible in the $J - W1 <$ 1.2 star counts ($<$ 0.1\% of the whole sample), we have used the $J - W$1 colour cut procedure for all the fields to separate the stars from the extra-galactic objects. Henceforth in the paper, we refer to GALEX and WISE+2MASS cross-matched sources with $J - W1 <$ 1.2 (GALEX+WISE+2MASS) as `UV-IR stars'.

\subsection{Photometric error cuts and completeness limits}

Figure 3 shows the distribution of UV-IR stars as a function of the GALEX UV magnitudes for AIS and MIS. Stars with NUV and FUV magnitude errors less than 0.5, 0.4, 0.3, 0.2 and 0.15 are displayed with magenta, green, red, blue and cyan colour lines respectively, whereas the black line represents the stars without any magnitude error cut. The typical $5\sigma$ magnitude limits of the NUV and FUV bands for AIS and MIS (see Section 2.1) are shown by vertical dashed lines. As seen from the histograms, a progressive stringent error cut eliminates the fainter stars. The completeness limits need to be established according to a given magnitude error. If we consider all stars without accounting for errors, the star counts go deeper but their values are not reliable due to the uncertainty on the magnitude measurement. This is particularly true for the FUV filter where some spurious detections can occur. Finally, we retained stars with magnitude error less than 0.2 in both bands as this error cut gives magnitude limits almost similar to the typical $5\sigma$ limits of the GALEX bands for AIS and MIS which are provided by \citet{Morrissey07}. 

We have applied magnitude error cuts (similar to the one shown in Figure 3 for UV-IR stars) in the original GALEX source catalog that includes all the GALEX detections and also in the matched GALEX source catalog obtained after cross-matching with the WISE+2MASS catalog. We find a loss of GALEX sources in the matched catalog when compared with the original GALEX source catalog at a specific magnitude error cut. The completeness limits for the original GALEX sources for NUV and FUV magnitude error cuts of 0.2 are 20.5/21.0 magnitude (FUV/NUV) in AIS and 22.5/22.5 magnitude (FUV/NUV) in MIS. The completeness limits at the same magnitude error cuts for the matched catalog become 20.0/20.5 magnitude (FUV/NUV) in AIS and 22.5/22.0 magnitude (FUV/NUV) in MIS and these limits are the same for the UV-IR stars too. For a specific magnitude error cut, the FUV and NUV completeness limits of the observed sources which are cross-matched to the surveys at longer wavelengths become brighter than the completeness limits of the unmatched GALEX source catalog due to the loss of faint sources in the former.

In order to examine which objects are affected by the limits of the combined surveys (GALEX+WISE+2MASS), we split the stars into two $NUV - W1$ colour intervals: hot ($NUV - W1 <$ 5) and cool ($NUV - W1 >$ 5) stars. We checked the completeness limit of the NUV band (AIS) for these two colour ranges. For hot stars, we found that the completeness limit of GALEX NUV (AIS) is reduced by 0.5 magnitude (i.e., the effective magnitude limit gets brighter). The GALEX completeness limit (AIS) of hot stars is therefore limited  by the depth of WISE, and similarly by the depth of 2MASS. For cool stars, the NUV (AIS) completeness limit is the same in the GALEX catalog alone and in the combined catalog with the near-IR surveys.

\section{Besan\c{c}on Galaxy model}

The Besan\c{c}on model is a population synthesis model based on a scenario of Galactic evolution and constrained by dynamics. In the model, five populations are taken into account: thin disc, thick disc, stellar halo, bar, and bulge \citep{Robin12}. The previous versions of the model are extensively described in \citet{Robin03}. We use the newest version of the model \citep[version of April 2013;][]{Robin12} which has been upgraded to include the FUV and NUV passbands of GALEX and the upcoming UVIT/ASTROSAT, by applying their filter transmission curves to produce UV star counts in various Galactic directions. The model uses a set of evolutionary tracks, a star formation rate and an IMF as defined in \citet{Haywood97}, to generate different stellar populations. The colours are computed from the Basel Stellar Library (BaSeL3.1) model atmospheres \citep{Westera02}. In this new version of the model, DA and DB type WDs are included using the evolutionary tracks and atmosphere models from \citet{Holberg06}. The luminosity functions are obtained assuming an initial-to-final-mass ratio ($ m_{I} = 9.1743m_{f}-3.6147$) from \citet{Kalirai08}. The distribution in age is computed assuming a lifetime on the main sequence (MS) from \citet{Eggleton89} and a lifetime on the giant branch of 15\% of the time on the MS. The repartition in DA (WD with hydrogen rich atmosphere) and DB (WD with helium rich atmosphere) is computed assuming that at T$_{\mathrm eff} >$ 40000 K they are all DA, and at  T$_{\mathrm eff} <$ 20000 K, 50\% are DB with a linear variation between 20000 K and 50000 K. The final luminosity function is normalised to fit \citet{Harris06}. Similarly, the BHBs are incorporated in the model by taking the evolutionary tracks from BaSTI (A Bag of Stellar Tracks and Isochrones) models \citep{Pietrinferni04}. Ultimately, the model produces UV star counts by Monte Carlo simulations using a set of stellar atmospheric models, observational photometric errors and extinction. 

The model incorporates an extinction (A$_{V}$) assuming an ellipsoidal distribution of diffuse absorbing matter, which follows an Einasto extinction law and is depicted by an adjustable normalization (extinction gradient) of 0.7 mag/kpc in the V band. We produced the model simulations towards various Galactic directions assuming the default extinction gradient. However, the default value of diffuse extinction (0.7 mag/kpc in the V band), which may not be appropriate at low latitudes, can be adjusted by adding a few absorbing clouds with a given adhoc distance and extinction from the \citet{Schlegel98} maps. This has been illustrated in Section 5.1. The ratios between UV band to visual extinction are taken to be 2.67 and 2.64 for the FUV and NUV band of GALEX, respectively, following the extinction law of \citet{Cardelli89}.

Stars in the simulated GALEX catalog have a UV colour, FUV and NUV magnitudes, a temperature range from 1650 K to 65000 K, log g from -1 to 9, all luminosity classes and a range of metallicities. In the simulations done for comparison with the GALEX observed star counts, the simulated stars are mostly MS stars ($\sim$77\%) with a small contribution from giants and subgiants ($\sim$17\%). The WDs are $\sim$6\% of the sample and reside at the bluer end of $FUV - NUV$ colour (see Sections 7).

\section{Comparison of the GALEX+WISE+2MASS and GALEX+SDSS stars}

Figure 4 shows the distribution of the GALEX AIS star counts (for field 5 in Table 1) as a function of NUV magnitude for the model simulation (solid line), GALEX+SDSS stars (dashed line), UV-IR stars (GALEX+WISE+2MASS: dashed-dotted line) and GALEX+WISE star counts with no 2MASS detection (dotted line). The error bars shown in the model star counts are due to Poisson noise. The NUV 5$\sigma$ detection limit \citep[NUV magnitude = 20.8;][]{Morrissey07} and the completeness limit ($\sim20.5$ magnitude; see Section 2.5) for AIS are demarcated by the solid and dashed vertical lines, respectively. Stars with NUV magnitude up to the completeness limit are well detected by the GALEX, SDSS, WISE and 2MASS surveys; a good agreement in star counts among the cross-matched surveys and the model simulations can clearly be seen in Figure 4.

 It is also evident from Figure 4 that the GALEX+SDSS stars are slightly more than the UV-IR stars in the NUV band at the fainter magnitudes. This discrepancy could be caused by 2MASS: since the time gap between the WISE and 2MASS surveys is $\sim$12 years, high proper motion stars may have moved outside the cross-matching radii.  However, stars with proper motions high enough to move by 3.0\arcsec\ in $\sim$12 years are very rare in a survey of a few square degrees. Another possibility is that the 2MASS J band, which has a 10$\sigma$ point source sensitivity limit of about 15.8 magnitude, does not penetrate deeply enough to provide counterparts for all WISE detections. Though GALEX+SDSS has a smaller sky area coverage and a fainter limit compared to GALEX+WISE+2MASS, both the selections yield a close match of the star counts at the brighter end. It is also possible that the GALEX+SDSS stars are still contaminated by faint galaxies and quasars. So, we preferred to use the star counts determined by the $J - W1$ colour cut (UV-IR stars) rather than the GALEX+SDSS stars.

\section{Data and model comparison}

We modelled the stellar density distribution of the Milky Way in UV using the Besan\c{c}on model of stellar population synthesis (as described in Section 3) for different regions of the sky. Four simulated catalogs for each of the fields chosen for our study were produced in order to reduce the statistical noise. Appropriate photometric errors were applied in the model to produce realistic simulations and the error information was assumed from the observed data which is a polynomial function of the magnitude. 

We can simulate the catalogs using the `small field' option which assumes that the density does not vary across the field, or using the `larger field' option with a given step in longitude and latitude, to account for the fields where the density can vary. We have used `small field' option for the small fields (e.g., area $<$ 15 deg$^{2}$) by providing the center $l$ and $b$ coordinates of the fields along with their coverage area. For the larger fields (e.g., area $>$ 15 deg$^{2}$), we have used the `larger field' option where we provide the range of $l$ and $b$ coordinates and a step size (e.g., from 1.0\degr\ to 2.5\degr\ for small to large fields, respectively) to cover the field. However, the gradients in the fields (Table 1) are small enough that considering either the center of the field or the range of $l$/$b$ does not make any difference in the predicted star counts.

\subsection{Comparison of observed UV star counts with the model in various fields}

Initially, simulations were performed for four GALEX tiles (fields 1 - 4 in Table 1), each covering an area of 0.785 deg$^{2}$ at the southern intermediate Galactic latitudes. We binned the model and the UV-IR stars in 0.5 magnitude intervals in the NUV band, for respective tiles of AIS and MIS, in the directions of the GC and the GAC. As shown in Figure 5, we found that the model star counts (solid line) match well the UV-IR stars (solid circles) as well as the GALEX+SDSS stars (open circles) up to the completeness limits of AIS (20.5 magnitude) and MIS (22.0 magnitude) for the regions at the southern intermediate Galactic latitudes.

In order to check the universal validity of the model, we produced simulated catalogs in various directions of the Galaxy covering a large area of the sky. Figures 6a and 6b show the comparison of the model-predicted NUV star counts (solid line) with the UV-IR stars (solid circles) as well as with the GALEX+SDSS stars (open circles) for AIS in the directions of the GC and the GAC, each covering 69.9 deg$^{2}$ area of the sky (fields 5 - 6 in Table 1). Similarly, Figures 6c and 6d represent the comparison of star counts for MIS in the directions of the GC and the GAC, covering an area of 22.77 deg$^{2}$ and 18.85 deg$^{2}$, respectively (fields 7 - 8 in Table 1). These fields are chosen at the northern intermediate Galactic latitude of the Galaxy. The error bars shown in the model-predicted star counts are due to Poisson noise. The maximum estimated asymmetric error in the observed counts is $\sim$2\% -- 10\% depending on the NUV magnitude bins (i.e., error increases towards the fainter magnitude bins), which is not shown in the plots. The model shows a good agreement with the observation (UV-IR stars and GALEX+SDSS stars) down to an NUV magnitude of $\sim$20.5 for AIS and 22.0 for MIS (see Figures 5 and 6).

We have also produced the model-predicted star counts for one of the passbands (NUVB4: 2505 - 2780 \AA, $\lambda_\mathrm{eff}$ = 2612 \AA) of upcoming UVIT/ASTROSAT which is shown by a dashed-dotted line. Star counts are enhanced in the UVIT NUVB4 band compared to the GALEX NUV band because NUVB4 covers a smaller wavelength range and its effective wavelength is longer than the effective wavelength of the NUV band. Most of the stars have flux peaks at longer wavelengths, such that $NUV - NUVB4$ is positive. Since the magnitudes are normalized to AB system, the integral of the filter does not matter while computing magnitudes, though narrower filters will demand longer exposure times to get the required magnitude.

The model-predicted star counts for the regions at the GHL, the GAR and the GP (solid line: fields 9 - 12 in Table 1) match well the UV-IR stars (solid circles) and the GALEX+SDSS stars (open circles) except the region at the GLL (see Figures 7a, 7b, 7c and 7d). As seen in Figure 7d, the model simulated NUV star counts (solid line) produced using the standard diffuse extinction do not match observation beyond NUV magnitude fainter than 18.5. This mismatch could be due to the default extinction gradient being used in the model not being sufficient at the GLL. We took the line of sight extinction (A$_\mathrm{V}$ = 0.1 magnitude) for the GLL from the \citet{Schlegel98} maps and then corrected the extinction by adding a cloud of A$_\mathrm{V}$ = 0.1 magnitude at a distance of 1 Kpc (Section 3). The model-predicted star counts after correcting the extinction (dashed line) show a good agreement with the UV-IR stars.

In Figure 8, we have shown the distribution in FUV magnitudes of the UV-IR stars (solid circles) and model-simulated (solid histograms) star counts for AIS and MIS (fields 5 - 8 in Table 1) towards the GC and the GAC. Despite the poor statistics, the model fit well the observations up to the completeness limit of the data sets (see Section 2.5). We have also produced the model-predicted star counts for one of the FUV passbands (BaF2: 1370 - 1750 \AA, $\lambda_\mathrm{eff}$ = 1504 \AA) of the forthcoming UVIT/ASTROSAT which is shown by a dashed-dotted line in Figure 8. Since the BaF2 passband range is close to the GALEX FUV passband, the UVIT model simulated FUV star counts match the GALEX observed FUV star counts reasonably well.

Overall the Besan\c{c}on model of stellar population synthesis upgraded to the UV passbands simulates star counts which are consistent with the observed GALEX star counts and can be used efficiently for the study of Galactic structure parameters.

\subsection{Latitude variation in star counts}

In order to study the latitude variation of UV star counts, we have chosen GALEX fields at 10\degr\ Galactic latitude intervals for $l \sim$ 50\degr. We determined NUV star counts per square degree in each field separately for the GALEX+SDSS stars and the UV-IR stars. As shown in Figure 9, the solid circles represent the UV-IR stars while the open circles show the GALEX+SDSS stars. The solid line represents the model generated star counts. The model errors due to Poisson noise are shown in the plot while the asymmetric errors on the UV-IR star counts which arise due to the propagation of photometric errors are not shown. The stellar density decreases from lower to higher Galactic latitudes in case of both observed and model star counts. The UV-IR star counts with NUV magnitude brighter than 20.5 magnitude match model simulations at the intermediate and high Galactic latitudes. However, a slight deviation of model simulated counts from observed counts is seen at low Galactic latitudes. This could be due to the default extinction gradient used in the model which might be inappropriate at low latitudes because some clouds can be present as discussed above for Figure 7(d).

\subsection{Comparison with the TRILEGAL model}

The predictions from the TRILEGAL model \citep{Girardi05}, which is another stellar population code, have been compared with UV star counts by \citet{Bianchi11a}. It was found that the TRILEGAL-predicted NUV star counts which show an overall good match to observations at brighter magnitudes are better at the northern high latitudes and the southern low latitudes. We produced NUV star counts using the 3 alternative IMFs that the TRILEGAL website\footnote{http://stev.oapd.inaf.it/cgi-bin/trilegal} proposes. However, we see in Figure 10 that the Besan\c{c}on model produces a better fit to real star counts than TRILEGAL does in the GHL field close to the pole as well as in the GAR field at the intermediate latitudes. Here we use a WD modeling similar to TRILEGAL, with small differences. The initial-to-final mass relation from \citet{Kalirai08} is used in the Besan\c{c}on model while TRILEGAL alternatively uses \citet{Marigo07} or \citet{Weidemann(00}, the latter giving a better fit to the GALEX data \citep[see Figure 9 in][]{Bianchi11a}. We also use different atmosphere models \citep{Holberg06}, while TRILEGAL uses either \citet{Koester08} or TLUSTY models \citep{Hubeny95}. \citet{Bianchi11a} pointed out that the difference between these two models is not larger than 0.05 magnitude in $FUV-NUV$ colour for most of the WDs. Finally, TRILEGAL does not consider DB WDs because it includes only WDs hotter than 18000 K, while we have taken them into account. However, the difference between TRILEGAL and the Besan\c{c}on model predictions is mainly due to the more detailed account for the settling of the disc with age in the Besan\c{c}on Galaxy model (the dynamical constraint which is used, forces the sub-components of the thin disc to follow a tied age/vertical scale height relation in agreement with the observed age/velocity dispersion relation).

\section{Distribution of the stars}

We find that the model reproduces the observed UV star counts as selected from the GALEX data. The star counts are dominated by MS stars, WDs and BHBs. The vertical distribution of different stellar populations depends on their structural parameters. In Figure 11a, we show the contribution of the thin disc (dotted line), thick disc (dashed line), halo (dashed-dotted line) and sum of the three populations (solid line) predicted by the model for an AIS field towards the GC at the intermediate Galactic latitude. The relatively bright stars are dominated by the thin disc at NUV magnitudes brighter than 18.5 whereas the thick disc and halo stars become significant at NUV magnitudes fainter than $\sim$18.5 and $\sim$19.5 respectively. This is very similar to the comparison made by \citet{Bianchi11a} for hot star candidates. Considering the stars with NUV magnitudes brighter than 20.5, we found that the thick disc stars are the most dominant population and $\sim$54\% - $\sim$60\% of the total population (depending on the Galactic direction).

We have shown the vertical distribution of the model simulated stars in Figure 11b. It is evident that the thin disc star counts (dotted line) dominate up to a distance of 1.5 kpc over the Galactic plane whereas the thick disc star counts (dashed line) dominate at distances between 1.5 and 4.0 kpc beyond which the halo stars (dashed-dotted line) dominate the total stellar population. A similar trend has been observed by both \citet{Du03} for BATC (Beijing-Arizona-Taiwan-Connecticut) multicolour photometric survey star counts and \citet{Phleps00} for CADIS (Calar Alto Deep Imaging Survey) deep star counts for regions at intermediate Galactic latitudes. 

\section{Blue hot stars}

$FUV - NUV$ colour is an important indicator of the spectral type of the stars. Particularly, UV colour can be used to identify hot BHBs and WDs \citep{Kinman07,Bianchi11a}, which emit most of their light in UV because of their high temperatures. The BHBs are comparatively more luminous in UV than the other population II stars. Similarly, the WDs which are the end product of the stellar evolution of the intermediate and low mass stars, provide important information about the Galactic disc star formation history. Comparing the observed $FUV - NUV$ colour of stars with the model, we were able to separate out the halo BHBs and disc WDs from the whole sample of stars. 

Figure 12 shows the comparison of GALEX $FUV - NUV$ colours for the UV-IR stars (solid circles) and model simulated star counts (solid-lined histogram) for the AIS fields towards the GC and the GAC. We have considered stars with NUV magnitude $<$ 20.5 and FUV magnitude $<$ 20.0 for the GALEX AIS survey. The $FUV - NUV$ colours of WD (dotted line) and BHB (long-dashed line) populations are also shown along with the UVIT $FUV - NUV$ (BaF2 - NUVB4) colour (dashed line) in the plot. Looking at the $FUV - NUV$ model predictions, the sources can be classified into two groups, the one with $FUV - NUV >$ 2.5 are the red cool stars and the other with $FUV - NUV <$ 2.5 are blue hot stars. The blue stars exhibit a bimodal distribution indicating the existence of two populations; the peak at $FUV - NUV \sim$ -0.5 are the hot WDs of the disc and the peak at $FUV - NUV \sim$ 2.0 are BHBs of the Galactic halo. In the Besan\c{c}on model, the temperature range of WDs is from 10000 K to 27000 K and that of BHBs is from 5000 K to 20000 K. Hotter stars with temperature greater than 27000 K are rare to be found in significant numbers in the data considered here.

The colour distributions in Figure 12 towards both the GC and the GAC show some differences between the model and observations. Specially, we notice that the very blue peak at $FUV-NUV <$ 0, due to hot WDs, is too high in the model. Moreover, there is a lack of stars in the GC field at 0 $< FUV - NUV <$ 1.5. In the colour range where the BHBs dominate, the number of predicted stars is well in agreement with the observations in both fields, indicating that the halo BHB density is well simulated. There is a dearth of model-simulated stars in the colour range, 2 $< FUV - NUV <$ 3.5, which is not understood yet and will be investigated in a further study. At $FUV - NUV >$ 4, the model lacks stars but more towards the GAC than towards the GC. This colour domain is mostly dominated by the thick disc MS stars. We guess that it is due to the scale length which will be investigated in a forthcoming paper.

Both, photometry and spectroscopy can be used to identify WDs and BHBs. Several large area sky surveys such as 2MASS, SDSS and GALEX have been used to distinguish them by appropriate colour selections and it is worth mentioning a few of the works. \citet{Kleinman13} produced the latest catalog of spectroscopically confirmed DA and DB type WDs from SDSS Data Release 7. Using the data from GALEX FUV and NUV imaging, \citet{Bianchi11b} presented a catalog of hot star candidates, particularly WDs. Similarly, the first selection of BHBs from SDSS colours was made by \citet{Yanny00} and then followed by many others \citep{Sirko04,Bell10,Deason11,Vickers12}. We have identified WD and BHB candidates using suitable GALEX $FUV - NUV$ colours. It was found from the model $FUV - NUV$ colour (Figure 11) that BHB and WD star candidates occupy the colour range, $1.5 < FUV - NUV < 2.5$ and  $FUV - NUV < 0.5$, respectively. In the mentioned colour range, we obtain a clean sample of WD candidates, whereas in the sample of BHB candidates, a contamination of non-BHB candidates, such as WDs and MS stars, constitute about 7\%. These colour ranges have been used for the separation of WD and BHB candidates from other populations in the observed sample. 

In order to substantiate our identification of the WD and BHB star candidates using GALEX $FUV - NUV$ colour, we compared them with their known 2MASS colours. $E(B - V)$ values for the stars were measured from \citet{Schlegel98} and converted to NUV, J and H extinction using \citet{Cardelli89} extinction law : $A(NUV) = 8.90E(B - V)$, $A(J) = 0.874E(B-V)$, and $A(H) = 0.589E(B-V)$. Figure 13a shows the $(J - K)_{o}$ versus $(NUV - J)_{o}$ colour-colour diagram for the BHB candidates. The sources at different latitude intervals are represented by different symbols. The dashed parallelogram encloses the area used by \citet{Kinman07} which contains 66\% of the BHB candidates selected on the basis of $FUV - NUV$ colour. Similarly, Figure 13b shows the $H - K$ versus $J - H$ colour-colour diagram for the WD candidates. The dashed rectangle encloses the area in the colour-colour diagram chosen from \citet{Hoard07} that contains a majority of the WD candidates of our sample. The location of our selected WD and BHB star candidates in the respective 2MASS colour window indicates that the $FUV - NUV$  colour can also be used as a potential tool in identifying WD and BHB candidates. This is a preliminary investigation and we will use this in our future work of an all sky study of these sources using the GALEX data.

\section{Conclusions}

The Besan\c{c}on model of stellar population synthesis has been previously checked at many different wavelengths from visible (U band) to mid-IR (12 $\mu$m). The model produces accurate star counts up to magnitude $\sim$ 22 in the visible or 18 in the K band. However, the stars that dominate the counts in the UV were not previously checked vis-a-vis model predictions. The availability of the GALEX data gives opportunity to check model predictions for high temperature, blue stars, specially BHBs from the halo and WDs from the disc. We have shown that the model performs very well for these types of stars as it does for other types. The model provides a good check that the population synthesis scheme gives predictions which are consistent with each other at all wavelengths. To do so, we make use of \citet{Holberg06} models which provide good stellar atmospheres and cooling tracks for WDs. However, the ratio between DA and DB type WDs has to be investigated more deeply. 

We have generally considered a simple dust distribution while limiting the comparisons to $|b| > 20\degr$. In future, we will compare the model at lower latitudes, in particular for the sake of analysis of the spiral structure, assuming the 3D extinction map from \citet{Marshall06}.

We also compared predictions in the UV bands from the TRILEGAL model with our model and found that the predictions of the Besan\c{c}on model are in better agreement with the observation than the TRILEGAL model as shown in Figure 10. However, in the faintest NUV magnitude bins TRILEGAL seems to be better in the GAR field. It will be something to look at carefully in the future using the all sky observations of GALEX and WISE, and we aim to present a detailed comparison between observations and the model.

We plan to complete the analysis by comparing model predictions with a variety of models of WDs, varying the tracks and investigating whether it could be possible to constrain the star formation history of the disc from the WDs distribution. Moreover, an analysis of the thick disc WD luminosity function could also be interesting for constraining the formation history of this old population, but it would require complementary kinematical data. We have seen that BHBs are a major component of GALEX stars. An analysis of this component could lead to constraints on the shape of the halo, once the contamination by extra-galactic objects is eliminated.

The final model can be safely used to predict star counts of various types in the UV wavelengths at the level of a few percent in many Galactic directions; the model produces star counts that match well down to FUV $\sim$ 20.0, NUV $\sim$ 20.5 for AIS, and FUV $\sim$ 22.5, NUV $\sim$ 22.0 for MIS. However, for the hot WDs, there is a mismatch of UV colours between the model and observation. A more detailed study is planned to explain the discrepancies by changing the WD luminosity function and the scale lengths alternatively. A study is also going on to better constrain the thick disc shape from large surveys in the visible and near-IR (Robin et al., in prep). We plan to further investigate the UV star counts with this revised model and the GALEX survey in the near future.

The Besan\c{c}on model is also developed to predict star counts in the UV passbands of the forthcoming UVIT telescope to be flown onboard ASTROSAT. We compared the model-predicted star counts at two of the UVIT filters with that of the GALEX observed star counts because of the similar wavelength coverage of both the instruments. The UVIT-predicted star counts are sensitively different from the GALEX observed star counts due to the differences in effective wavelengths. UVIT star counts will be very useful to separate out different stellar populations since they have several UV colours and better angular resolution compared to GALEX, which in turn will help us to estimate the structural parameters of the Galaxy with better precision.

\begin{acknowledgements}

The authors thank the anonymous referee for useful comments and suggestions that improved the content of the paper.

GALEX (Galaxy Evolution Explorer) is a NASA small explorer launched in 2003 April. We gratefully acknowledge NASA's support for construction, operation, and science analysis for the GALEX mission, developed in cooperation with the Centre National d'Etudes Spatiales of France and the Korean Ministry of Science and Technology. This work has made use of the data products from the Wide-field Infrared Survey Explorer (WISE), Two Micron All Sky Survey (2MASS) and Sloan Digital Sky Survey (SDSS). We also thank the UVIT/ASTROSAT team for providing the UVIT filter response curves.

Simulations were executed on computers from the Utinam Institute of the Universit\'{e} de Franche-Comt\'{e}, supported by the R\'{e}gion de Franche-Comt\'{e} and Institut des Sciences de l'Univers (INSU). We acknowledge the support of the French ``Agence Nationale de la Recherche'' under contract ANR-2010-BLAN-0508-01OTP. Many thanks to Bernard Debray who is responsible for providing the web interface for the Besan\c{c}on Galaxy model.

This work was fully or partially supported by the Gaia Research for European Astronomy Training (GREAT-ITN) Marie Curie network, funded through the European Union Seventh Framework Programme (FP7/2007-2013) under grant agreement No. 264895. J.J.V. is a fellow at the International Max Planck Research School for Astronomy and Cosmic Physics at the University of Heidelberg.

\end{acknowledgements}

\newpage

\begin{table*}[t]
\caption{Details of the GALEX fields. The areas of different fields are chosen depending on the availability of GALEX, WISE+2MASS and SDSS overlapping regions.
\label{table1}}
\begin{center}
\begin{tabular}{ccccccccc}

\hline

Field & Longitude & Latitude & Survey type & Area  & Location & Number of tiles & NUV exposure & FUV exposure\\
& range (deg) & range (deg) & & (deg$^{2}$) &  & & time (sec) & time (sec) \\
\hline
\hline
1  &  47.79      &   -43.56   & AIS  &   0.785 & GC  & 1  & 175         & 175\\
2  &  129.33     &   -43.15   & AIS  &   0.785 & GAC & 1  & 258         & 258\\
3  &  47.01      &  -42.65    & MIS  &   0.785 & GC  & 1  & 1589        & 1589\\
4  &  146.57     &   -46.51   & MIS  &   0.785 & GAC & 1  & 1657        & 1657\\
5  &  1 - 15     &  50 - 60   & AIS  &   69.9  & GC  & 89 & 64 - 400    & 64 - 272\\
6  &  160 - 175  &  50 - 60   & AIS  &   69.9  & GAC & 89 & 90 - 442    & 96 - 220\\
7  &  13 - 29    &  35 - 41   & MIS  &   22.77 & GC  & 29 & 1541 - 2176 & 1541 - 2176\\
8  &  154 - 162  &  38 - 44   & MIS  &   18.85 & GAC & 24 & 1597 - 4457 & 1512 - 3066\\
9  &  230 - 240  &  42 - 50   & AIS  &   46.34 & GAR & 59 & 80 - 438    & 61 - 231\\
10 &  40 -50     &  75 - 85   & AIS  &   14.92 & GHL & 19 & 62 - 292    & 62 - 249\\
11 &  30 - 32    &  21 - 29   & AIS  &   13.35 & GLL & 17 & 105 - 421   & 105 - 230\\
12 &  0 - 20     &  84 - 88   & AIS  &   4.71  & GP  & 6  & 107 - 383   & 107 - 271\\
13 &  49 - 51    &  17 - 23   & AIS  &   7.07  &  -   & 9  & 130 - 199   & 130 - 199\\
14 &  48 - 53    &  29 - 33   & AIS  &   11.0  &  -   & 14 & 79 - 184    & 79 - 184\\
15 &  46 - 53    &  37 - 43   & AIS  &   20.42  &  -   & 28 & 96 - 384    & 95 - 265\\
16 &  46 - 54    &  46 - 54   & AIS  &   17.28 &  -   & 22 & 94 - 345    & 73 - 226\\
17 &  46 - 54    &  57 - 63   & AIS  &   11.78 &  -   & 15 & 152 - 342   &  152 - 229\\
18 &  46 - 54    &  67 - 73   & AIS  &   7.07  &  -   & 9  & 90 - 340    &  90 - 169\\
19 &  44 - 54    &  77 - 83   & AIS  &   4.71  &  -   & 6  & 75 - 259    &  75 - 151\\  
\hline
\end{tabular}
\end{center}
\end{table*}

\newpage
\begin{figure*}
\centering
\subfloat{
 \includegraphics[width=7.5cm]{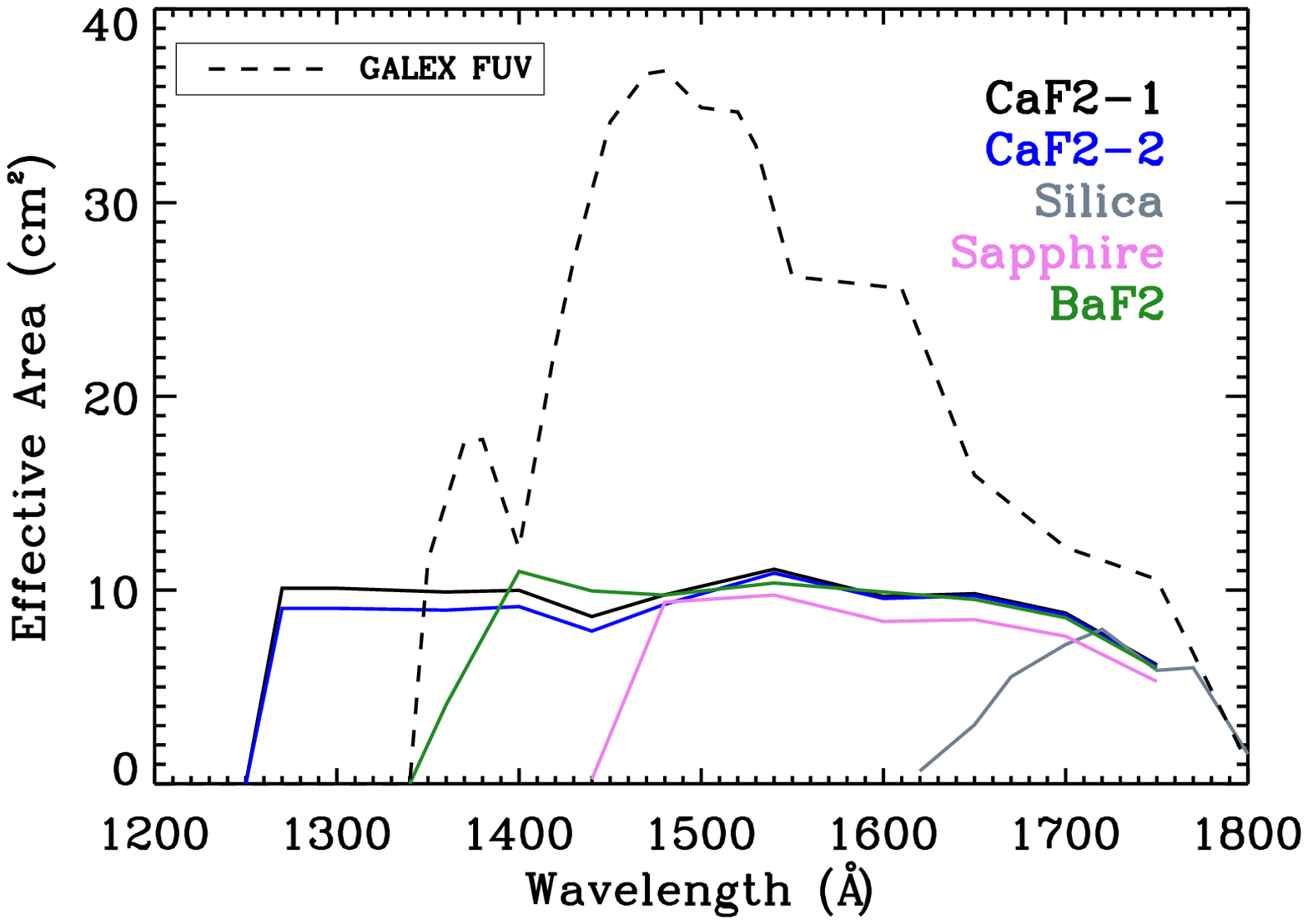}}
\subfloat{
 \includegraphics[width=7.5cm]{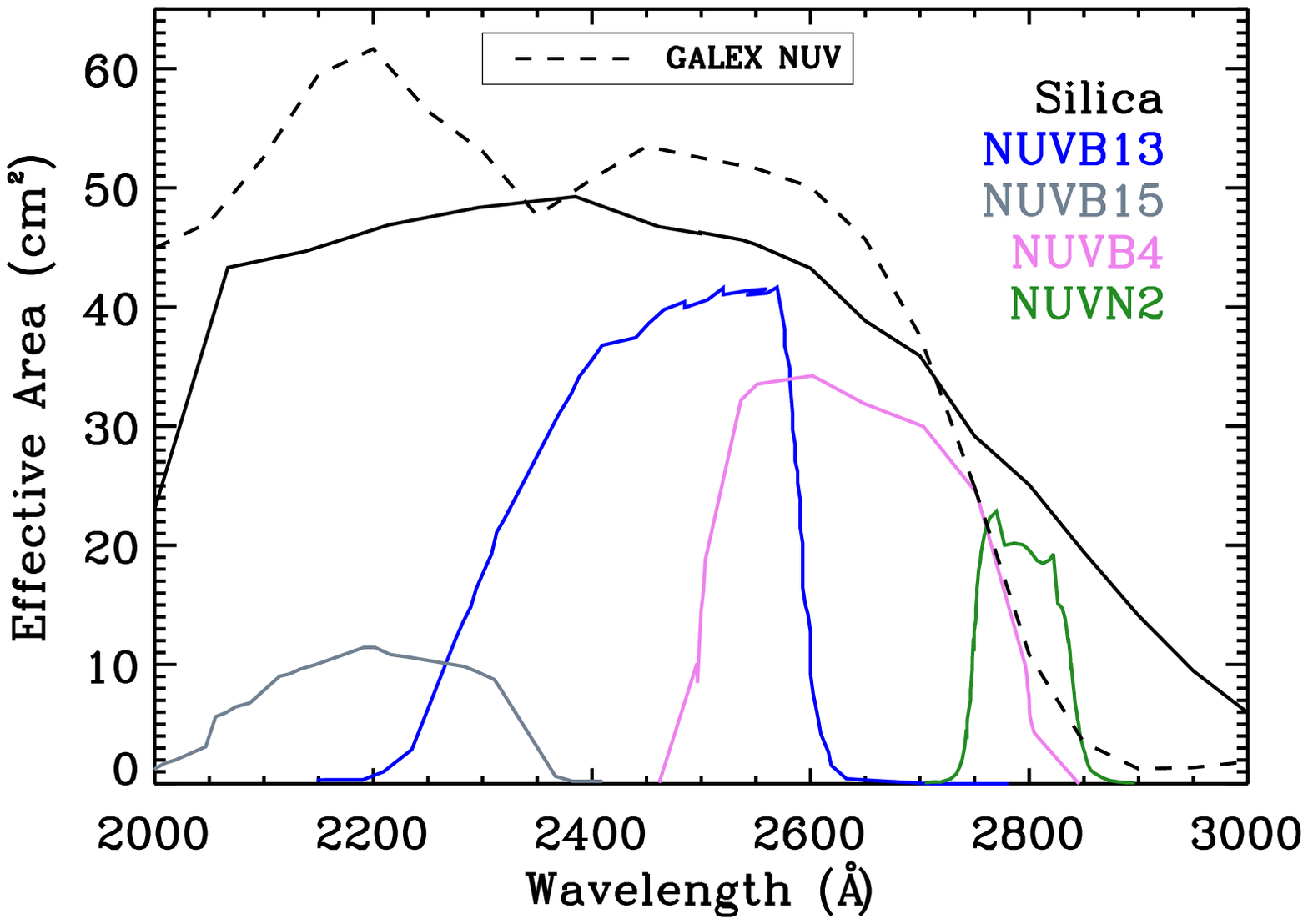}}
\caption{Effective area versus wavelength plots for the GALEX FUV and NUV bands are shown in relation to the FUV and NUV filters of UVIT/ASTROSAT for imaging mode. The left panel shows the five FUV filters of UVIT/ASTROSAT in different colours along with the GALEX FUV filter (dashed line) and the right panel shows the same for the NUV filters.  
\label{fig1}
}
\end{figure*}

\begin{figure*}
\centering
\subfloat{
 \includegraphics[width=7.5cm]{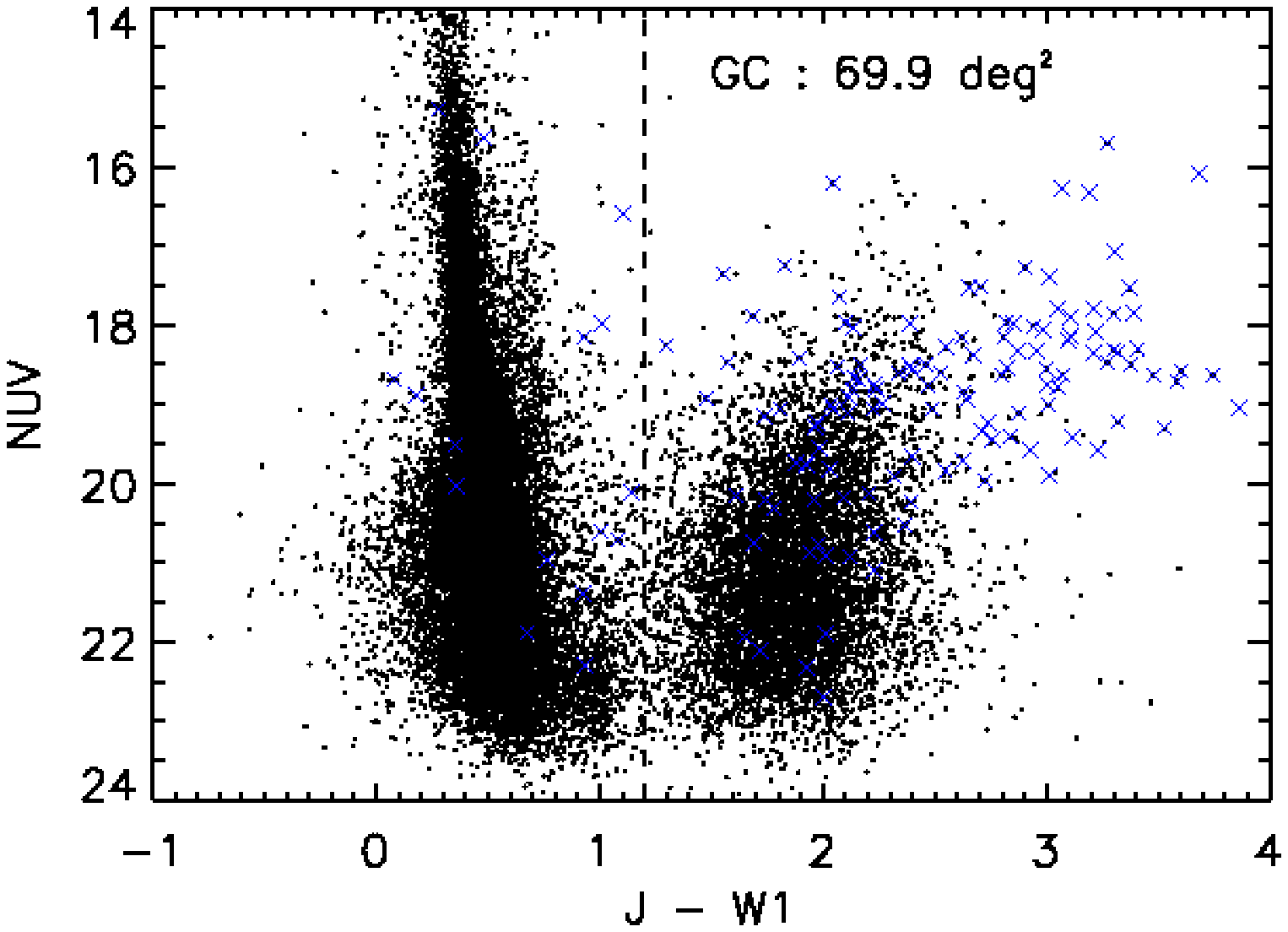}}
\subfloat{
 \includegraphics[width=7.5cm]{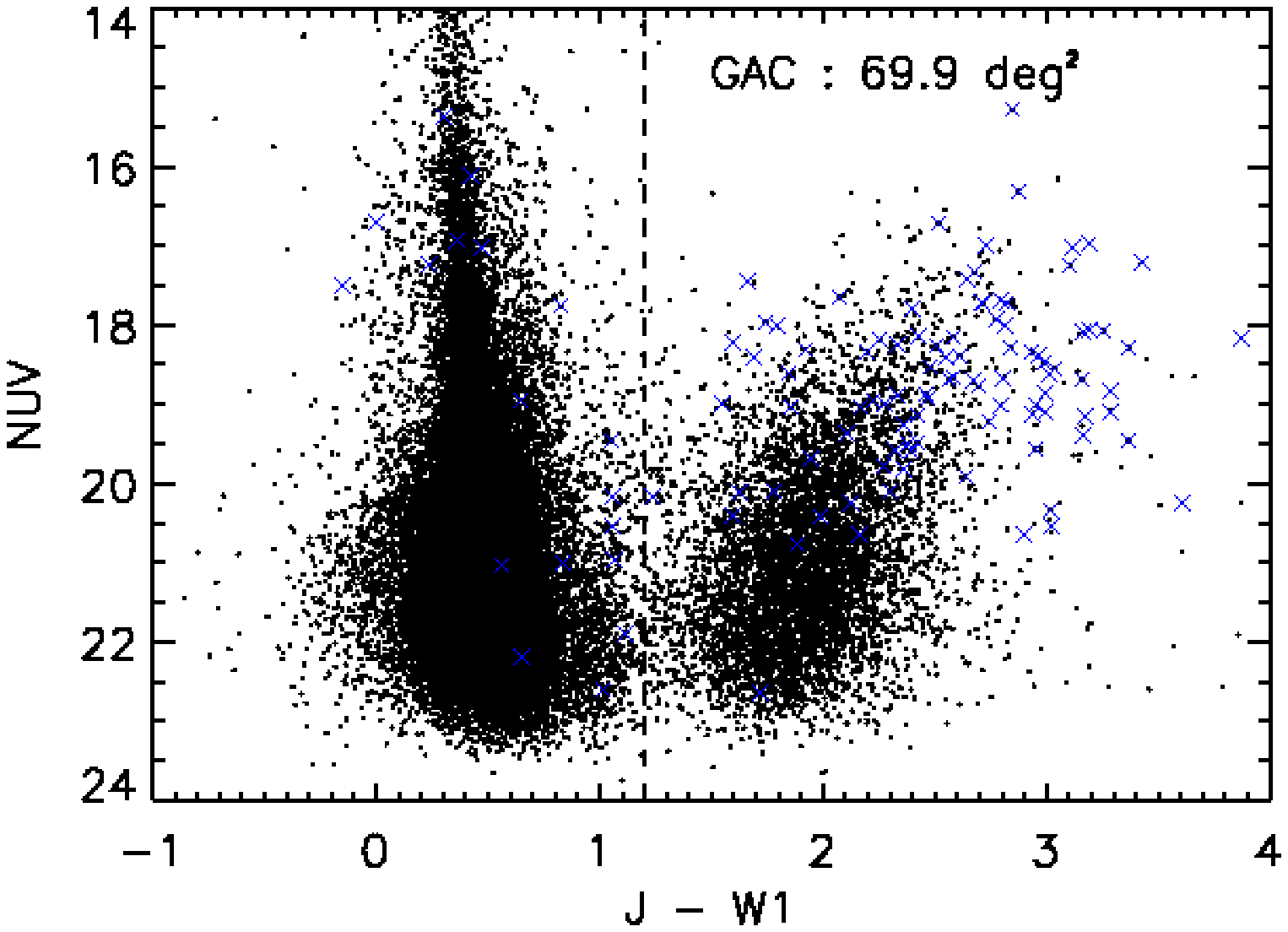}}
\caption{The diagrams show $J$ (2MASS) $-$ $W1$ (WISE) versus NUV CMD for the GALEX and WISE+2MASS cross-matched sources for the AIS fields towards the GC and the GAC. The matched sources are clearly separated in two groups indicating isolation of stars ($J - W1 <$ 1.2) from the extra-galactic sources ($J - W1 >$ 1.2). The vertical dashed line shows the limit that we choose for selecting the point sources ($J - W1 <$ 1.2). QSOs are shown by blue crossed symbols (see the text).
\label{fig2}
}
\end{figure*}

\begin{figure*}
\centering
 \includegraphics[width=14cm]{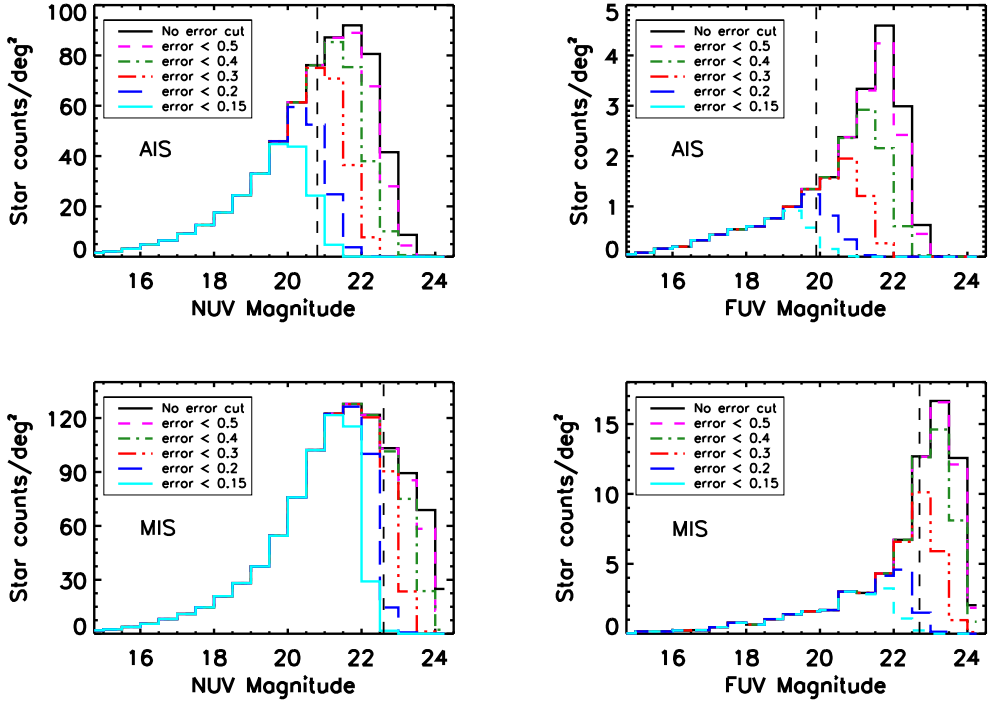}
\caption{Distribution of the UV-IR stars as a function of the GALEX NUV and FUV magnitudes for AIS and MIS. The stellar sources obtained after applying various magnitude error cuts are shown by different line styles in colours. The vertical dashed lines represent the respective 5$\sigma$ detection limits of the GALEX bands for typical exposure times as mentioned in Section 2.
\label{fig3}}
\end{figure*}

\begin{figure*}
\centering
 \includegraphics[width=12.cm]{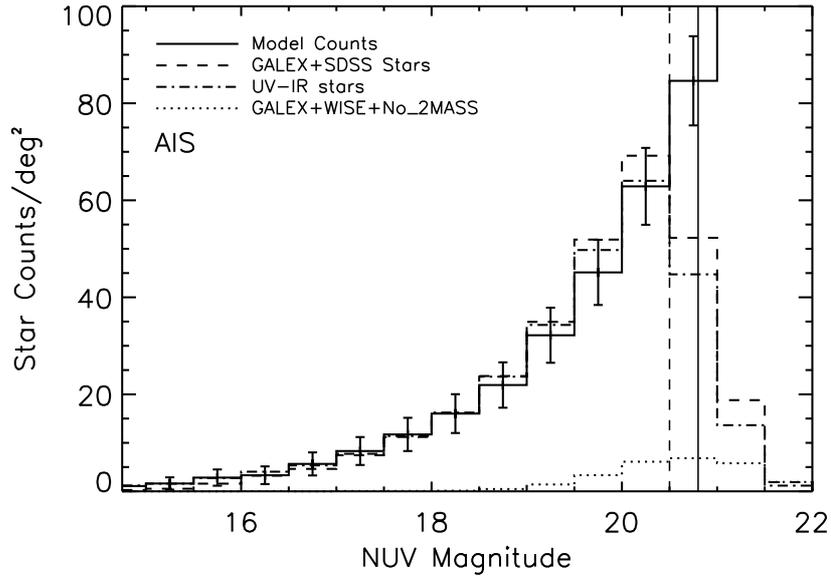}

\caption{Distribution of the UV-IR stars (GALEX+WISE+2MASS : dashed-dotted line), GALEX+SDSS stars (dashed line) and model-simulated star counts (solid line) for the AIS field towards the GC covering 69.9 deg$^{2}$ of the sky (field 5 in Table 1). The dotted line represents the GALEX+WISE sources with no 2MASS counterparts. The star counts are binned in 0.5 magnitude interval in NUV magnitude. The error bars in the model-simulated star counts are due to Poisson noise. The NUV 5$\sigma$ detection limit (NUV magnitude = 20.8) is shown by a solid vertical line. The UV-IR star counts show a turn over at NUV magnitude $\sim$ 20.5 (demarcated by a dashed vertical line).
\label{fig4}}
\end{figure*}

\begin{figure*}
\centering
\subfloat[]{
 \includegraphics[width=7.5cm]{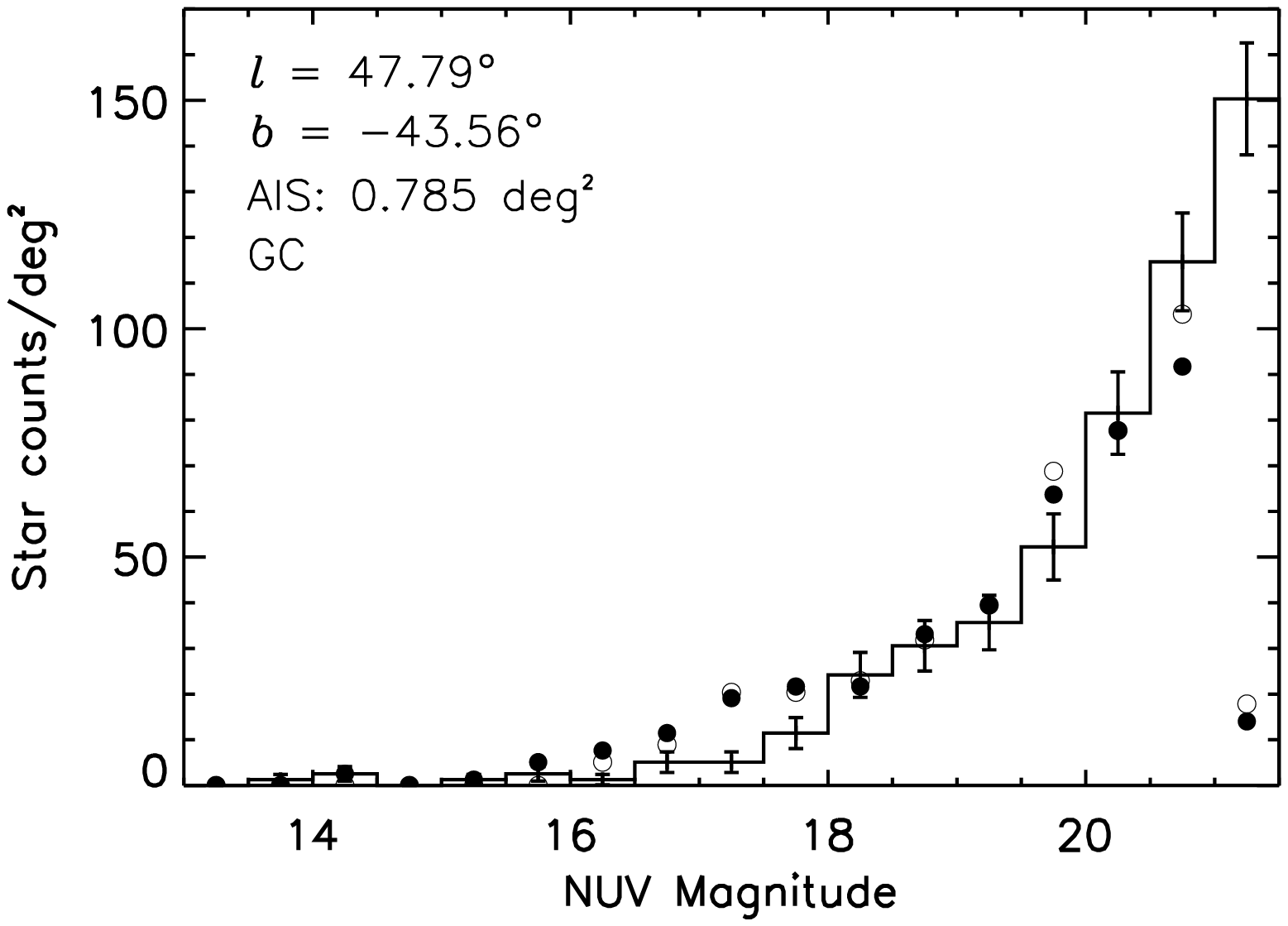}}
\subfloat[]{
 \includegraphics[width=7.5cm]{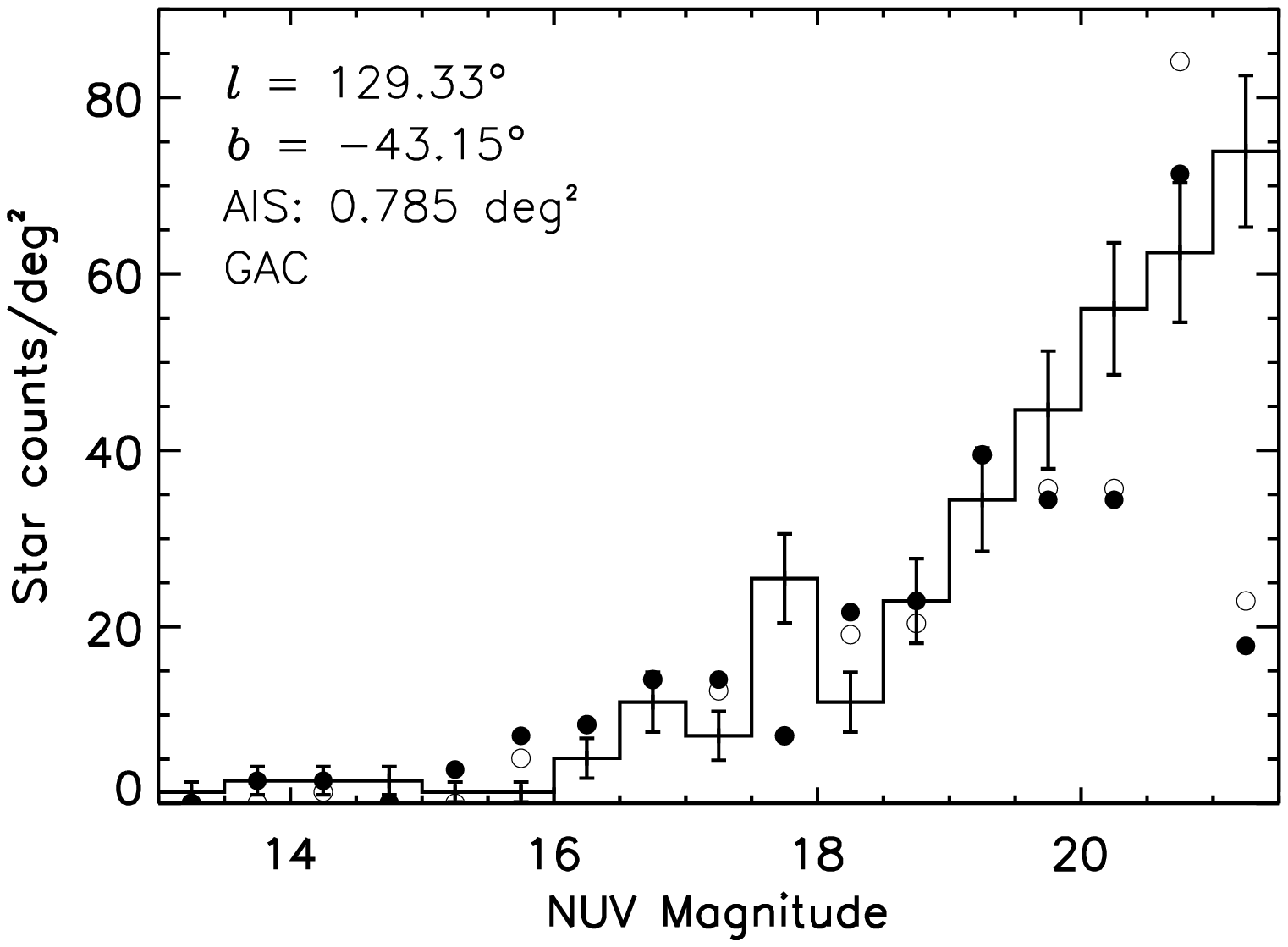}} \\
\subfloat[]{
 \includegraphics[width=7.5cm]{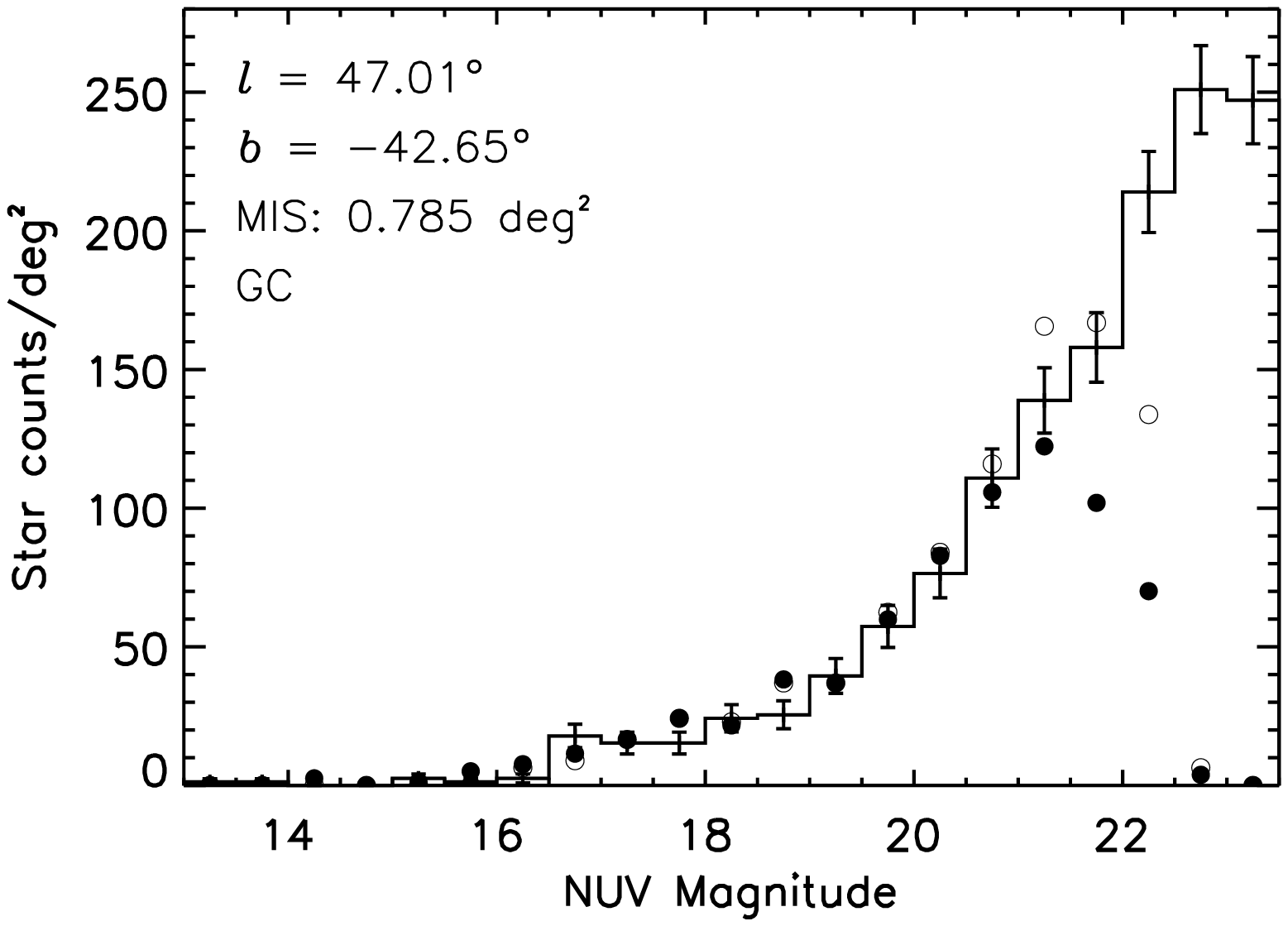}}
\subfloat[]{
\includegraphics[width=7.5cm]{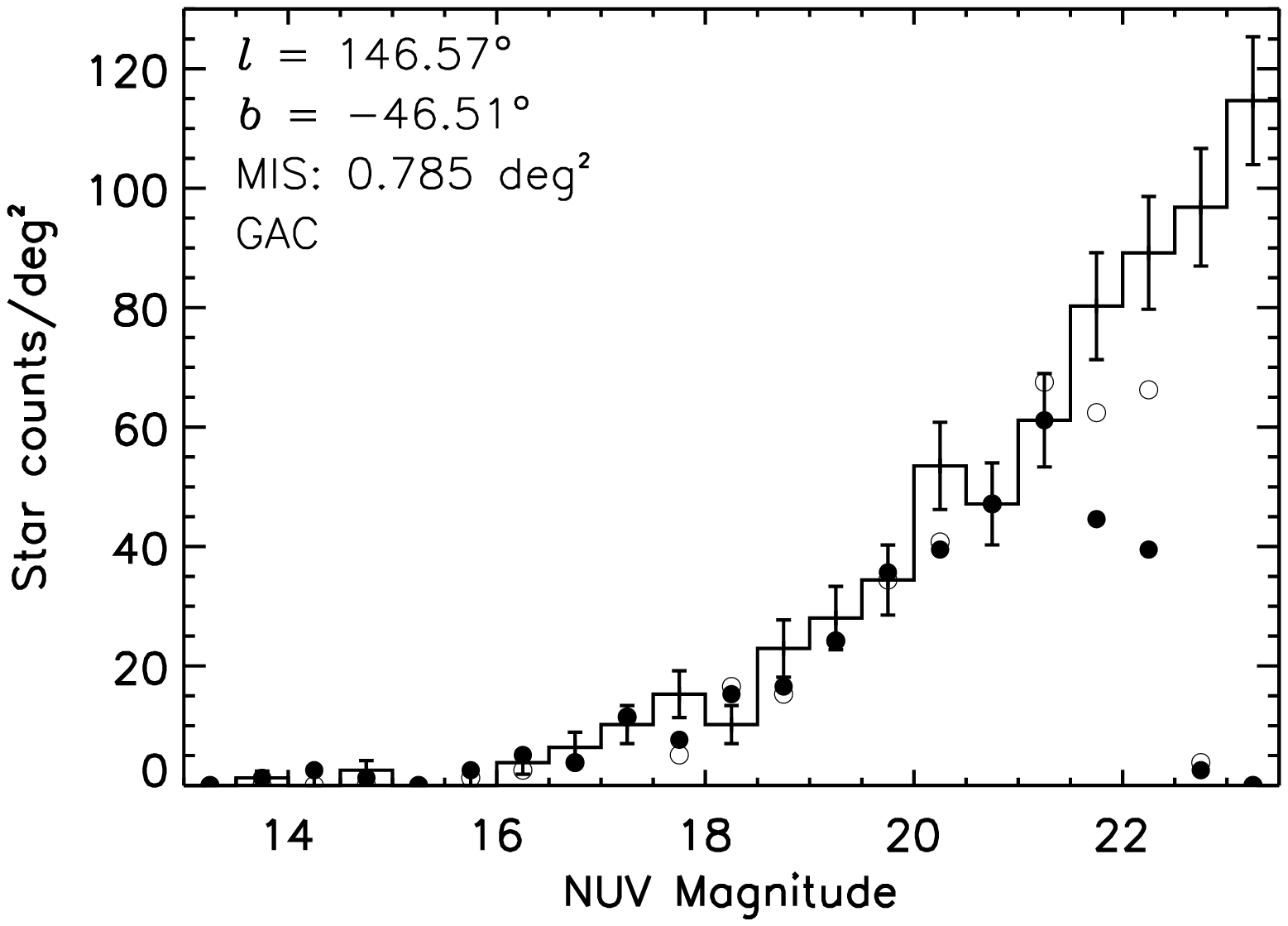}}
\caption{Comparison of the UV-IR stars (solid circles) with model-predictions (solid line) as a function of NUV magnitudes for the regions at the southern intermediate Galactic latitudes. The open circles represent the GALEX+SDSS stars. The plots are for the fields towards the GC and the GAC for individual GALEX AIS and MIS tiles, each covering an area of 0.785 deg$^{2}$ (fields 1-4 in Table 1). The error bars shown in the model star counts are due to Poisson noise, while the asymmetric errors in the observed star counts are not shown in the plot. 
\label{fig5}
}
\end{figure*}

 \begin{figure*}
 \centering
\subfloat[]{
 \includegraphics[width=7.5cm]{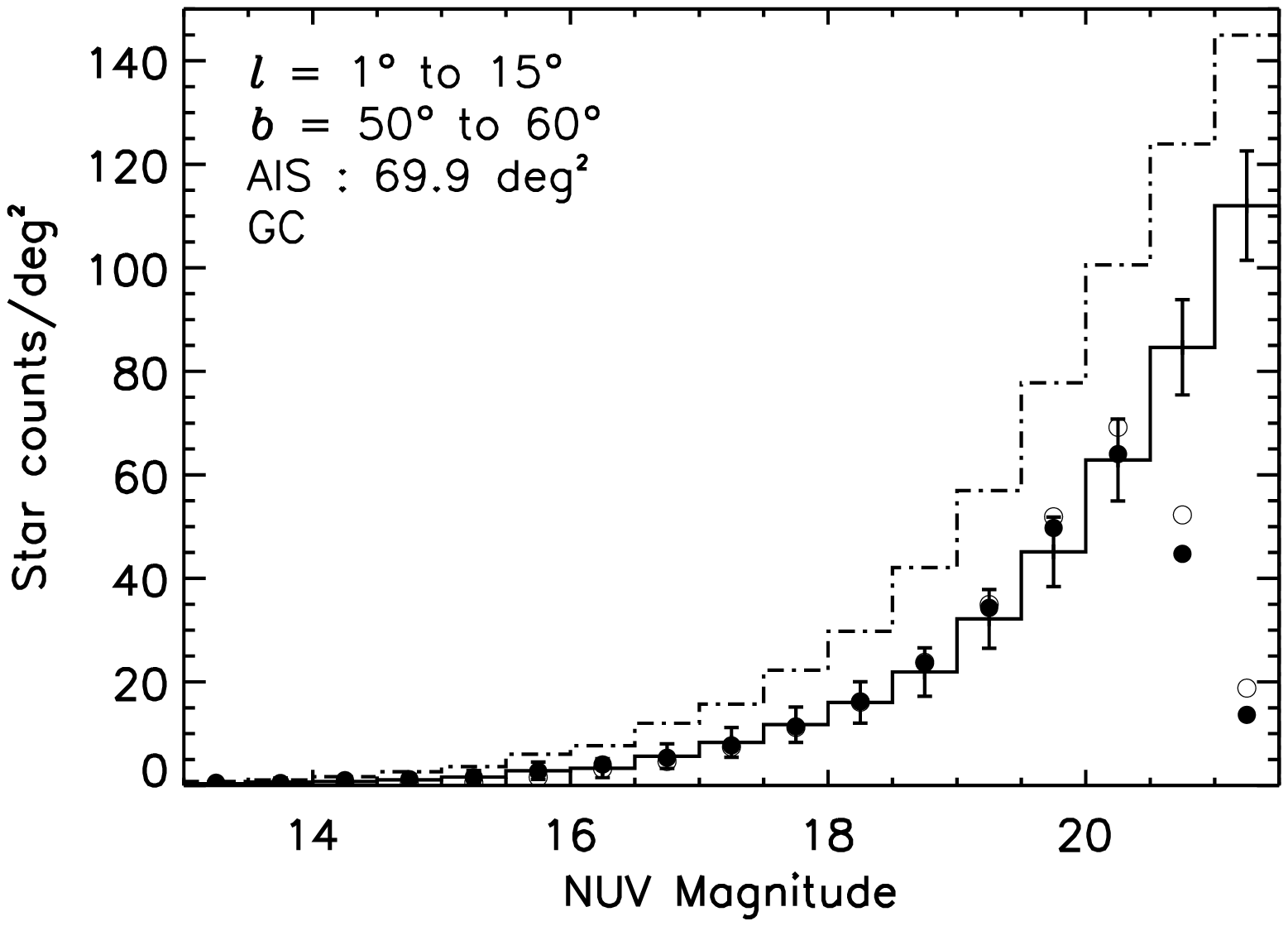}}
\subfloat[]{
 \includegraphics[width=7.5cm]{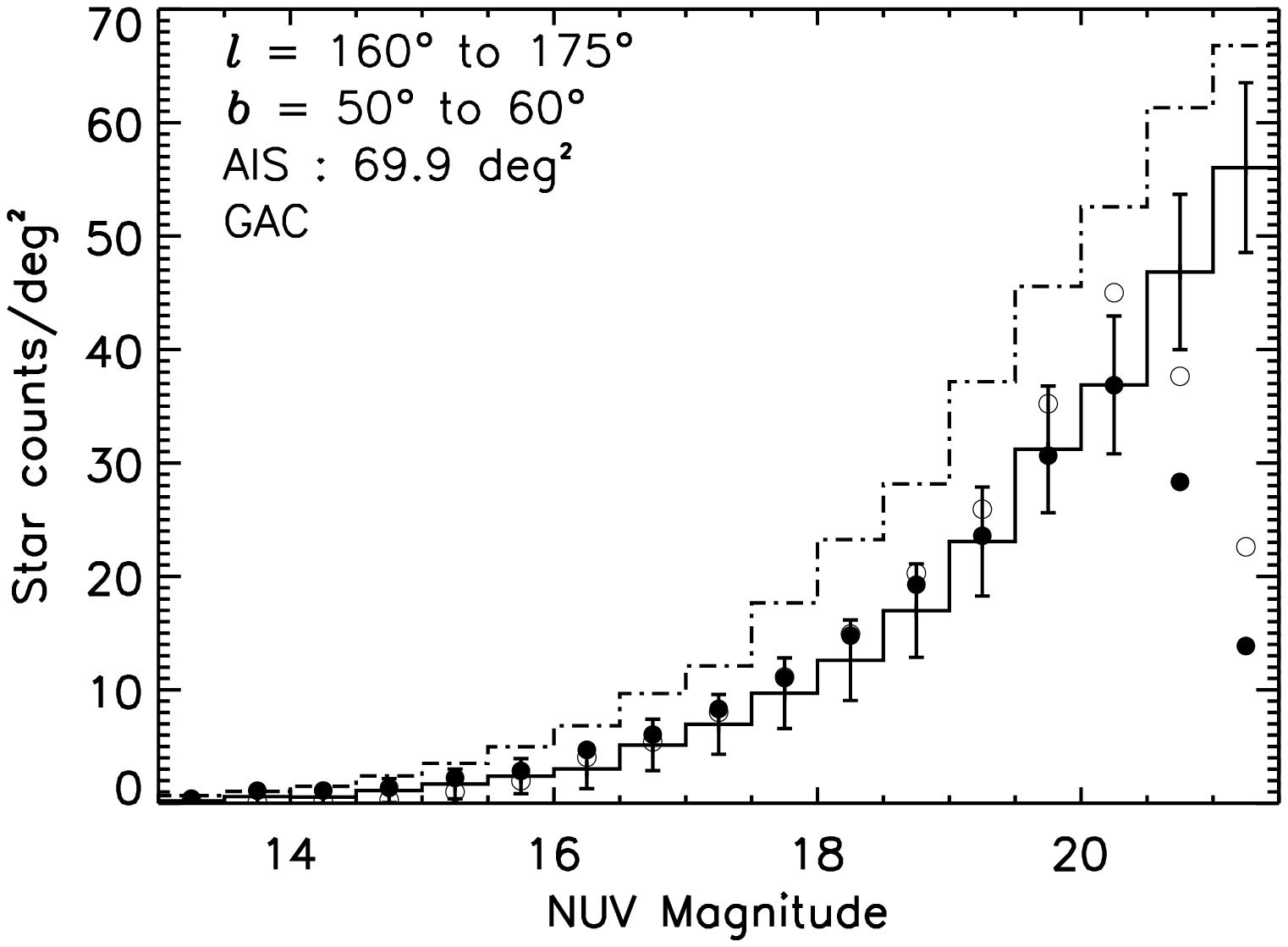}} \\
\subfloat[]{
 \includegraphics[width=7.5cm]{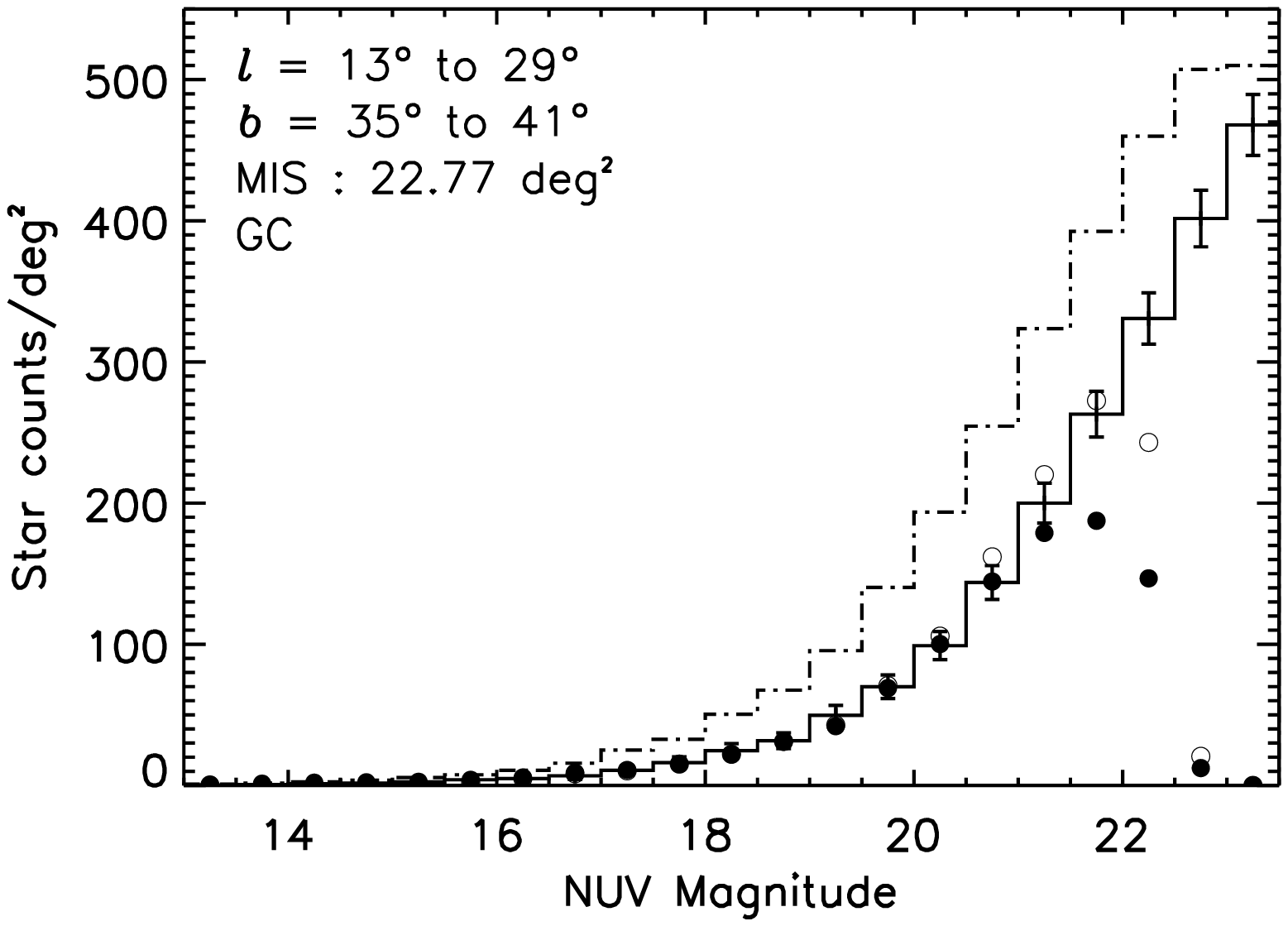}}
\subfloat[]{
\includegraphics[width=7.5cm]{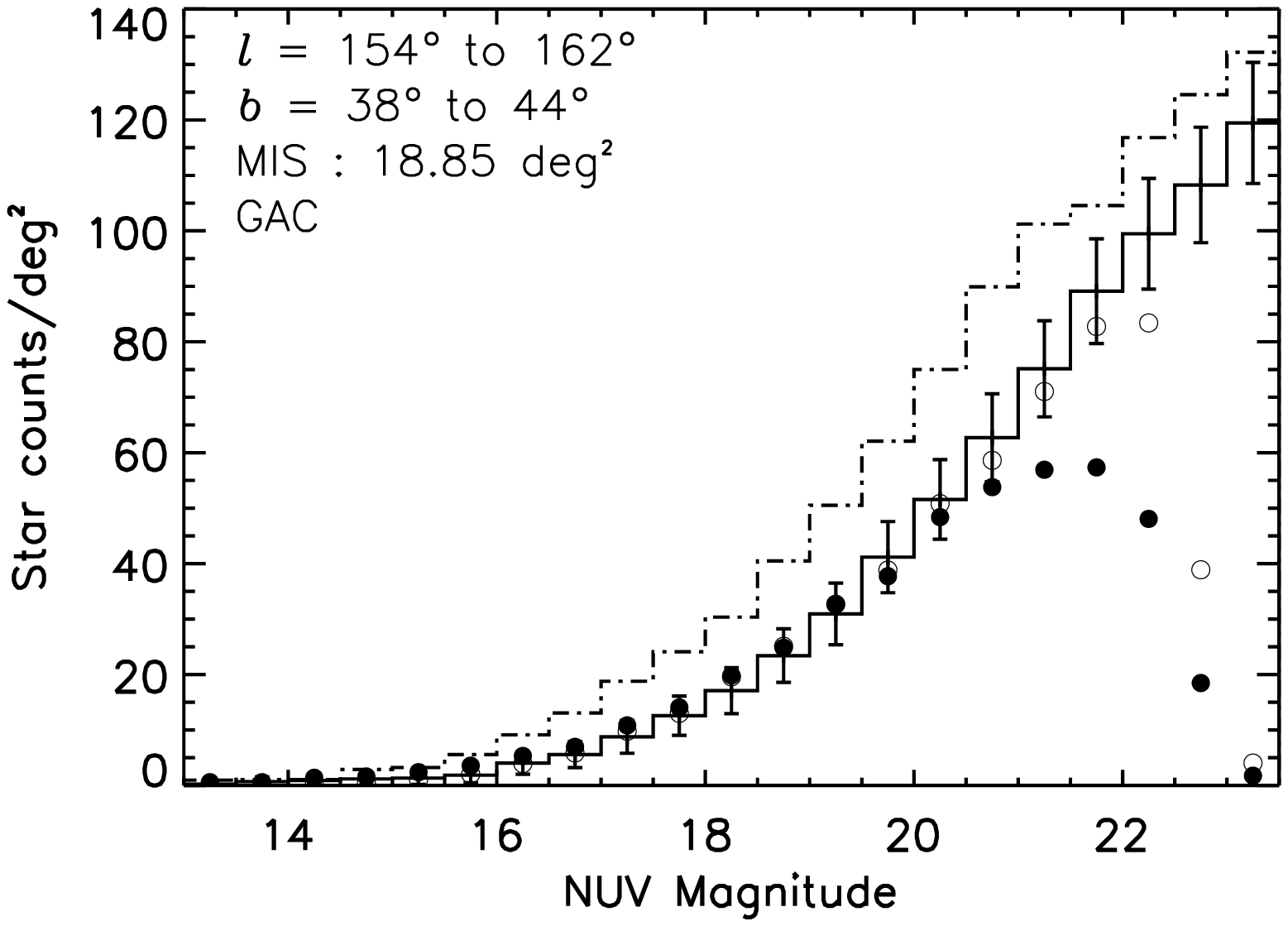}}\\
\caption{Comparison of the model-predicted star counts (solid line) with the UV-IR stars (solid circles) as well as with the GALEX+SDSS stars (open circles) for the GALEX fields at the northern intermediate Galactic latitudes. The Galactic coordinate ranges, survey types and area coverages are mentioned in each panel. The dashed-dotted line shows the model-simulated star counts for the NUVB4 band of UVIT/ASTROSAT (2505 - 2780 \AA, $\lambda_\mathrm{eff}$ = 2612 \AA). The error bars shown in the model counts are due to Poisson noise.
\label{fig6}}
\end{figure*}

\begin{figure*}
\centering
\subfloat[]{
\includegraphics[width=7.5cm]{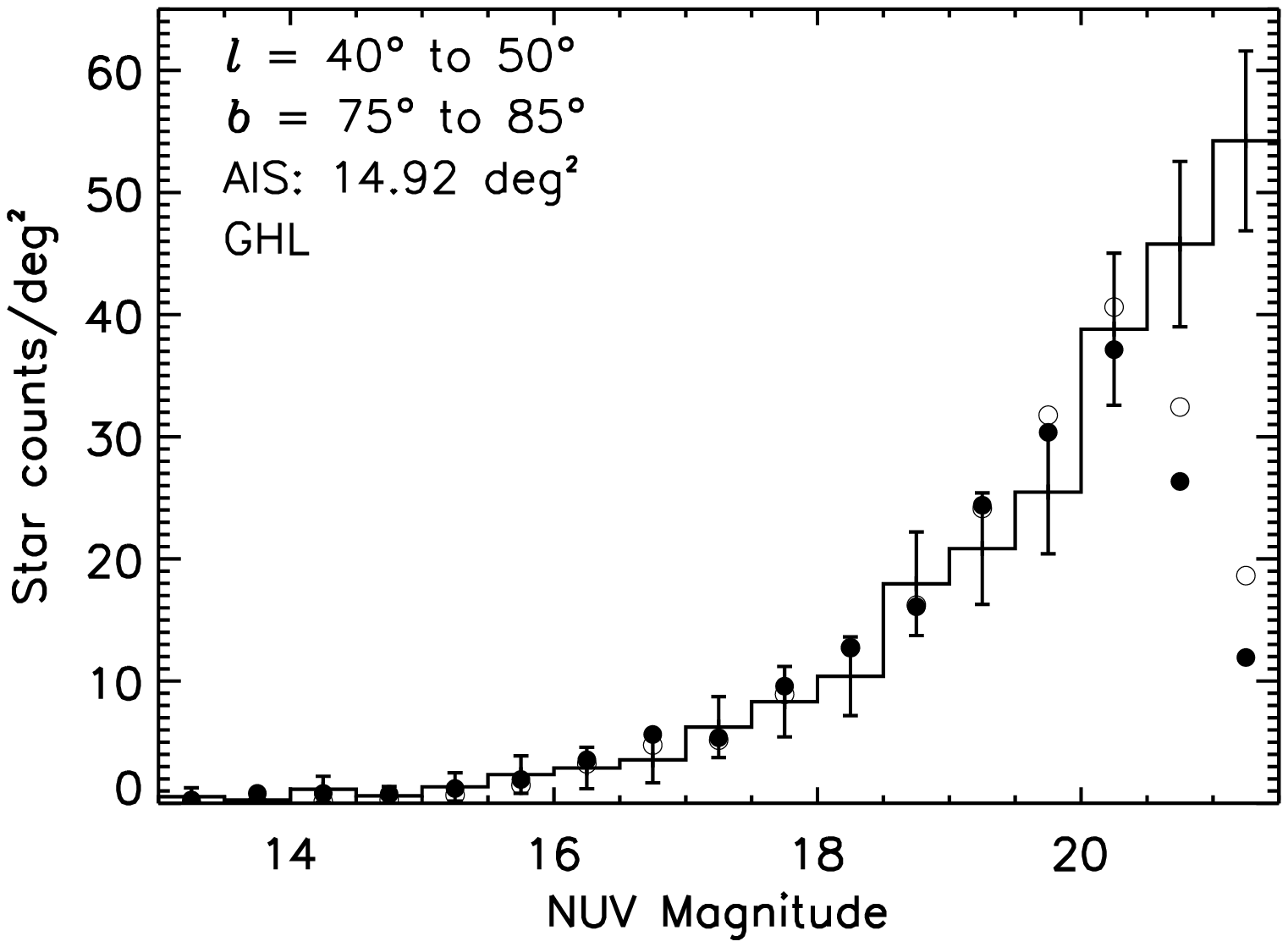}}
\subfloat[]{
 \includegraphics[width=7.5cm]{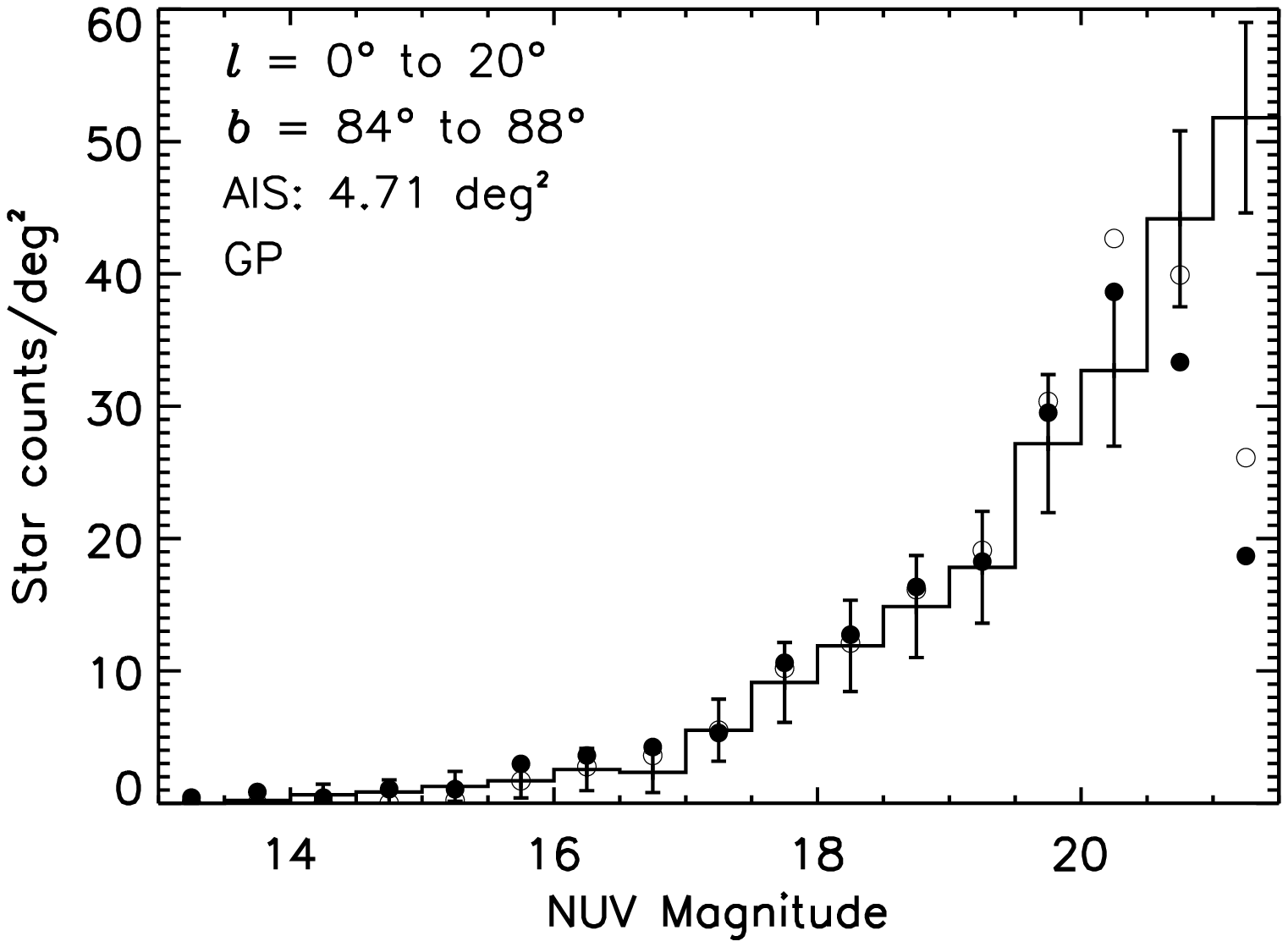}}\\
\subfloat[]{
 \includegraphics[width=7.5cm]{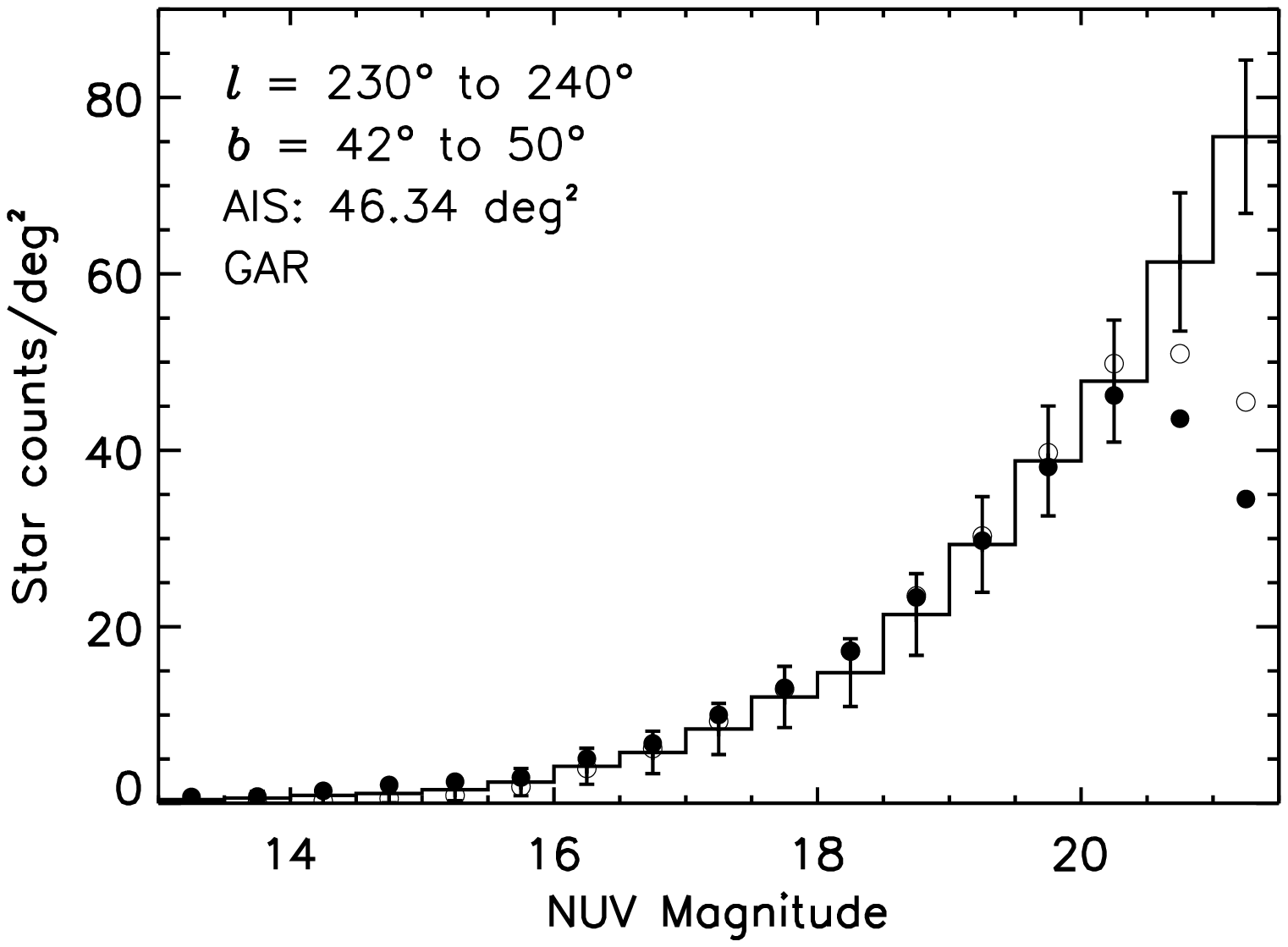}}
\subfloat[]{
\includegraphics[width=7.5cm]{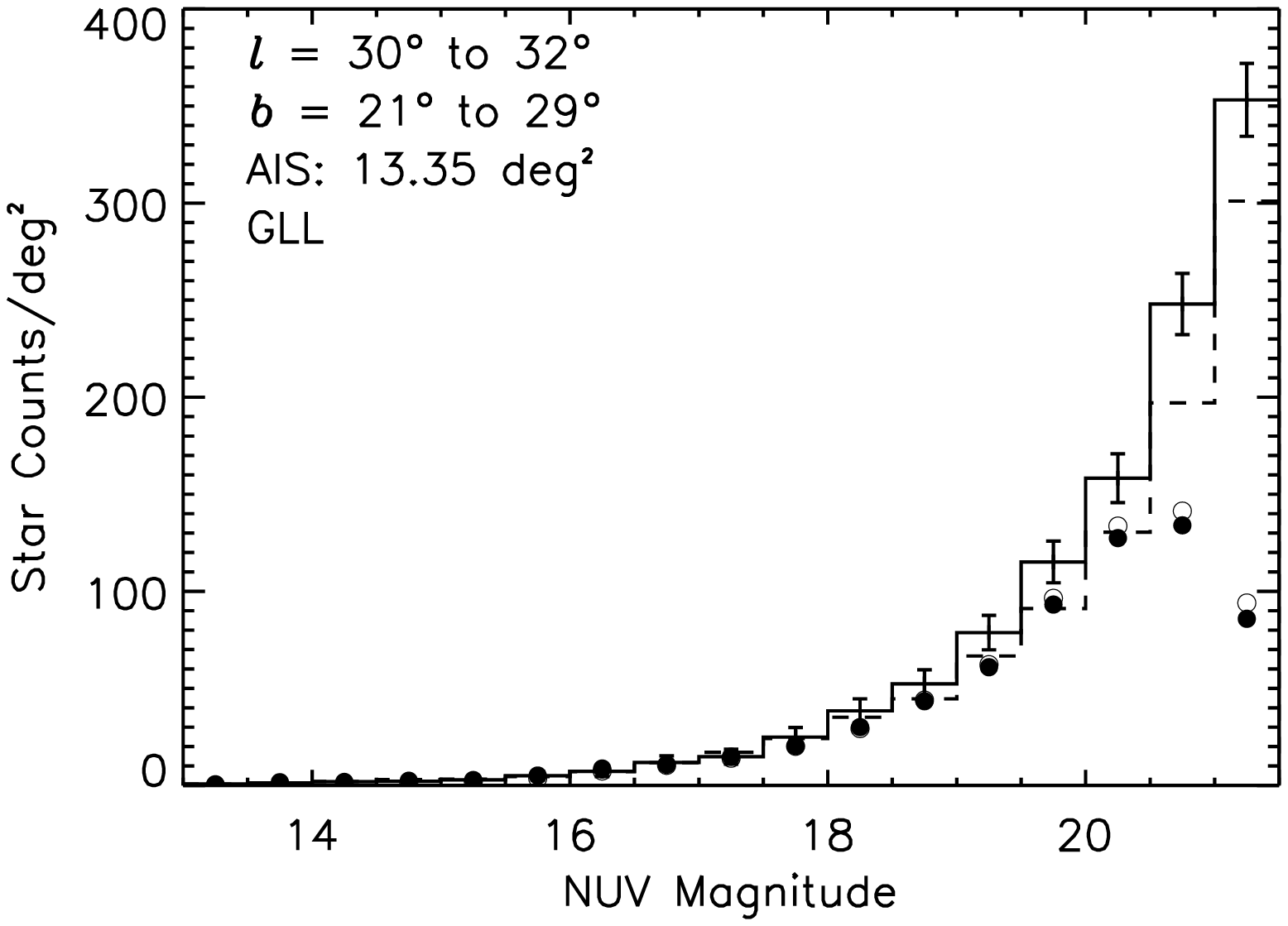}}\\
\caption{Comparison of model-predictions (solid line) with the UV-IR stars (solid circles) and GALEX+SDSS stars (open circles) as a function of the GALEX NUV magnitude for the regions at the GHL (7a), the GP (7b), the GAR (7c) and the GLL (7d). In plot 7d, the solid line represents the model NUV star counts produced assuming the standard diffuse extinction (as in other fields) whereas the dashed line is the same after correcting the extinction using the value from the \citet{Schlegel98} maps (see Section 5). The model error bars shown in the plots are due to Poisson noise.
\label{fig7}}
\end{figure*}

\begin{figure*}
\centering
\subfloat{
 \includegraphics[width=7.5cm]{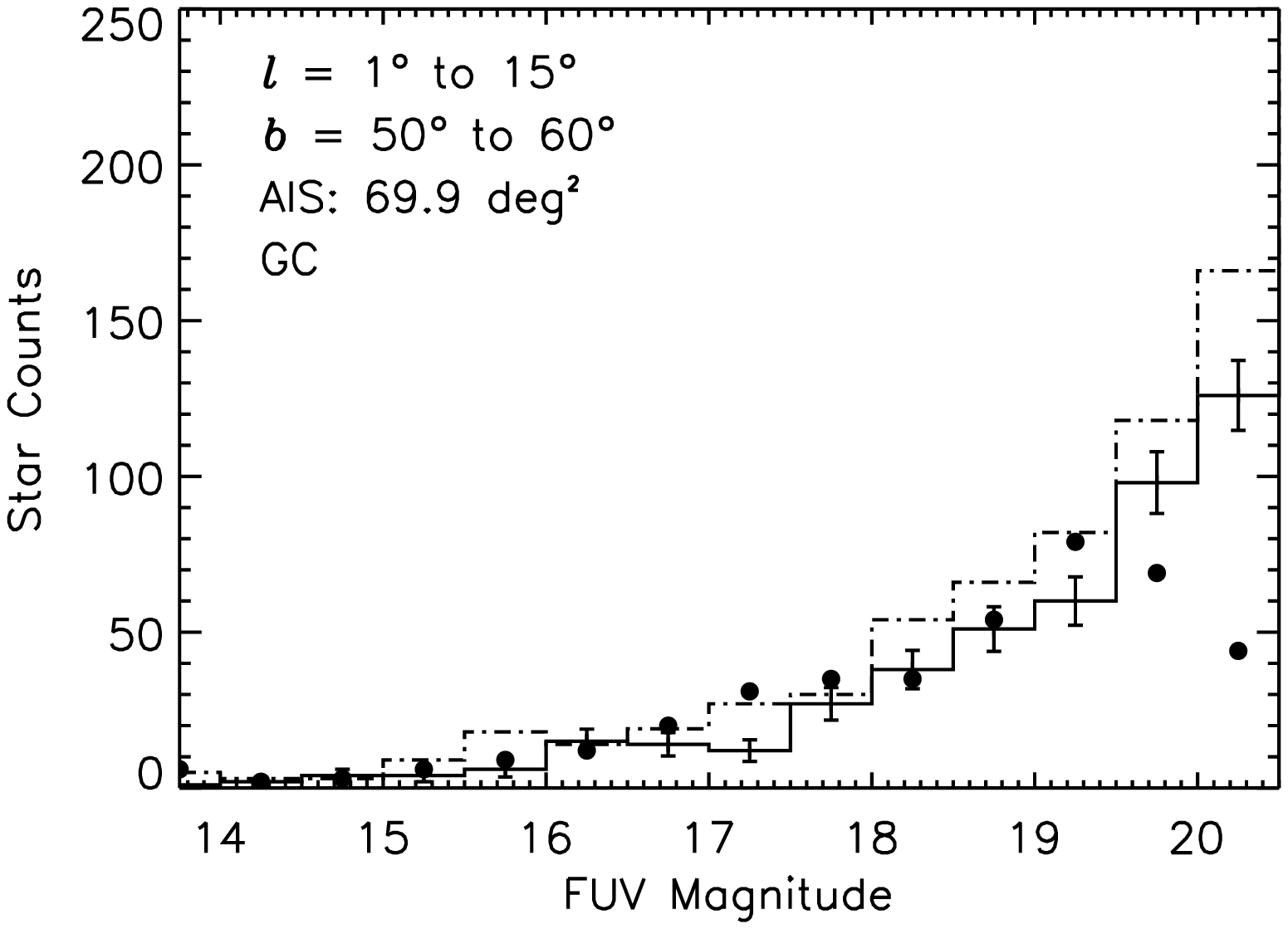}}
\subfloat{
 \includegraphics[width=7.5cm]{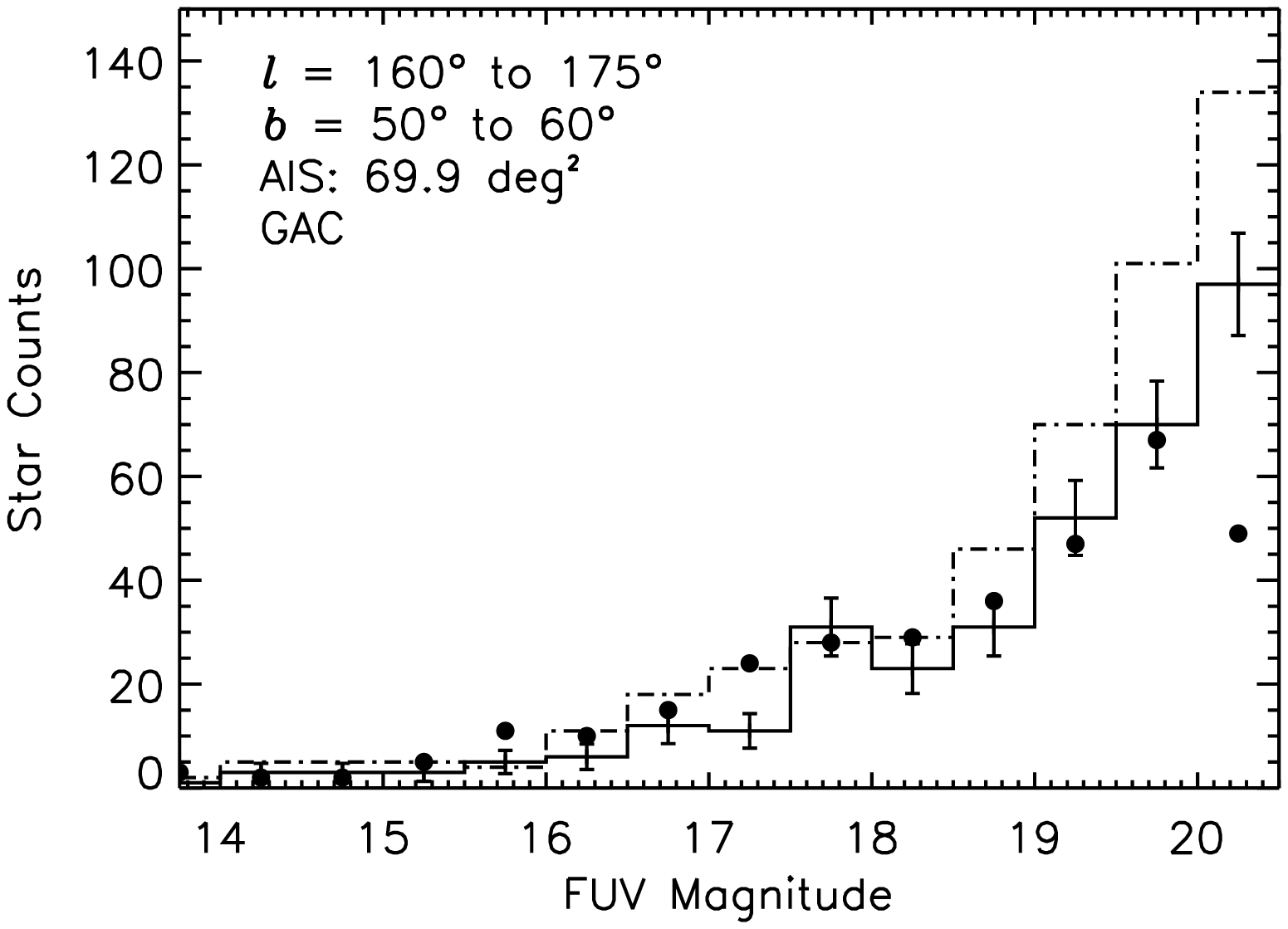}}\\
\subfloat{
 \includegraphics[width=7.5cm]{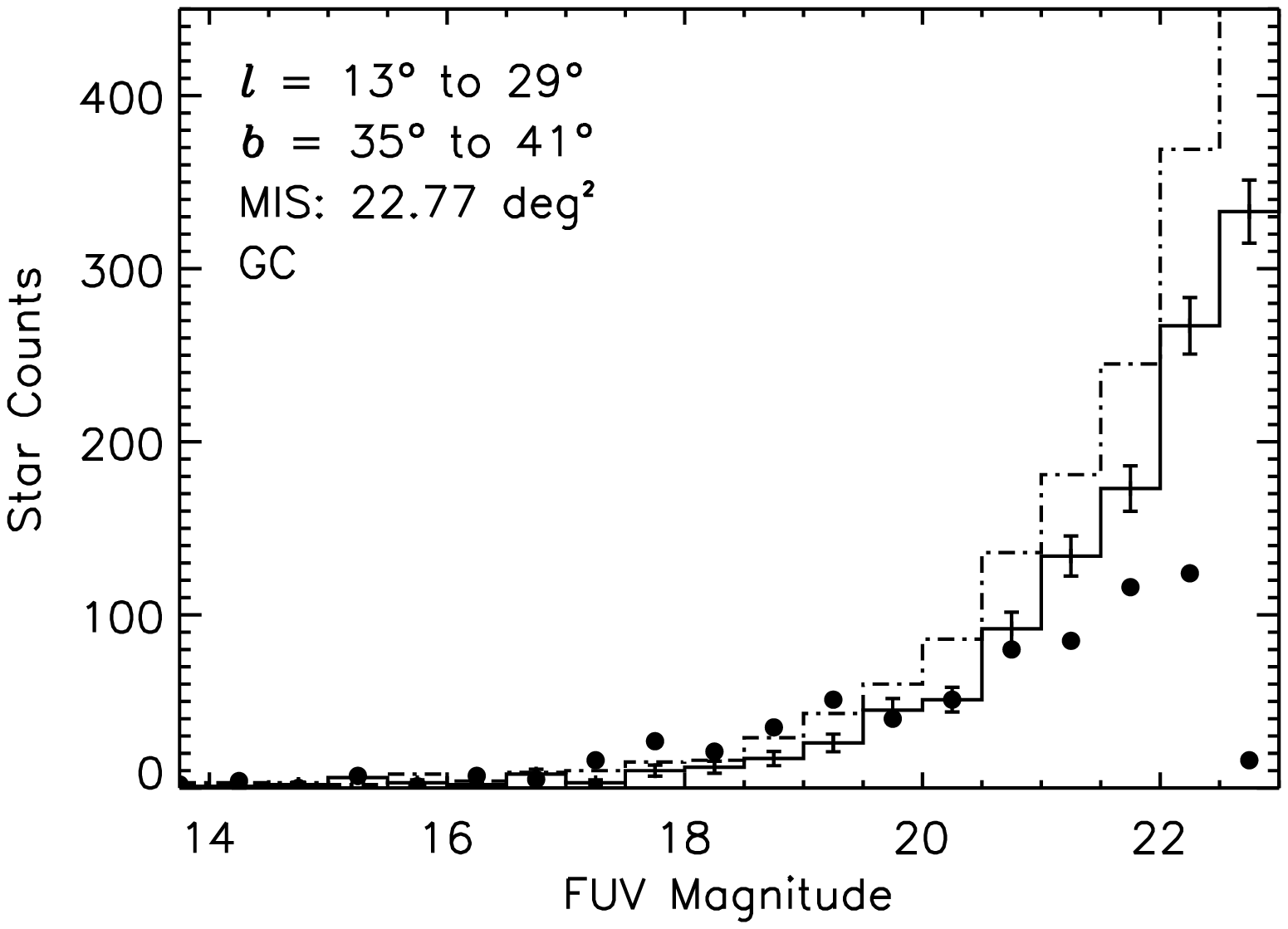}}
\subfloat{
 \includegraphics[width=7.5cm]{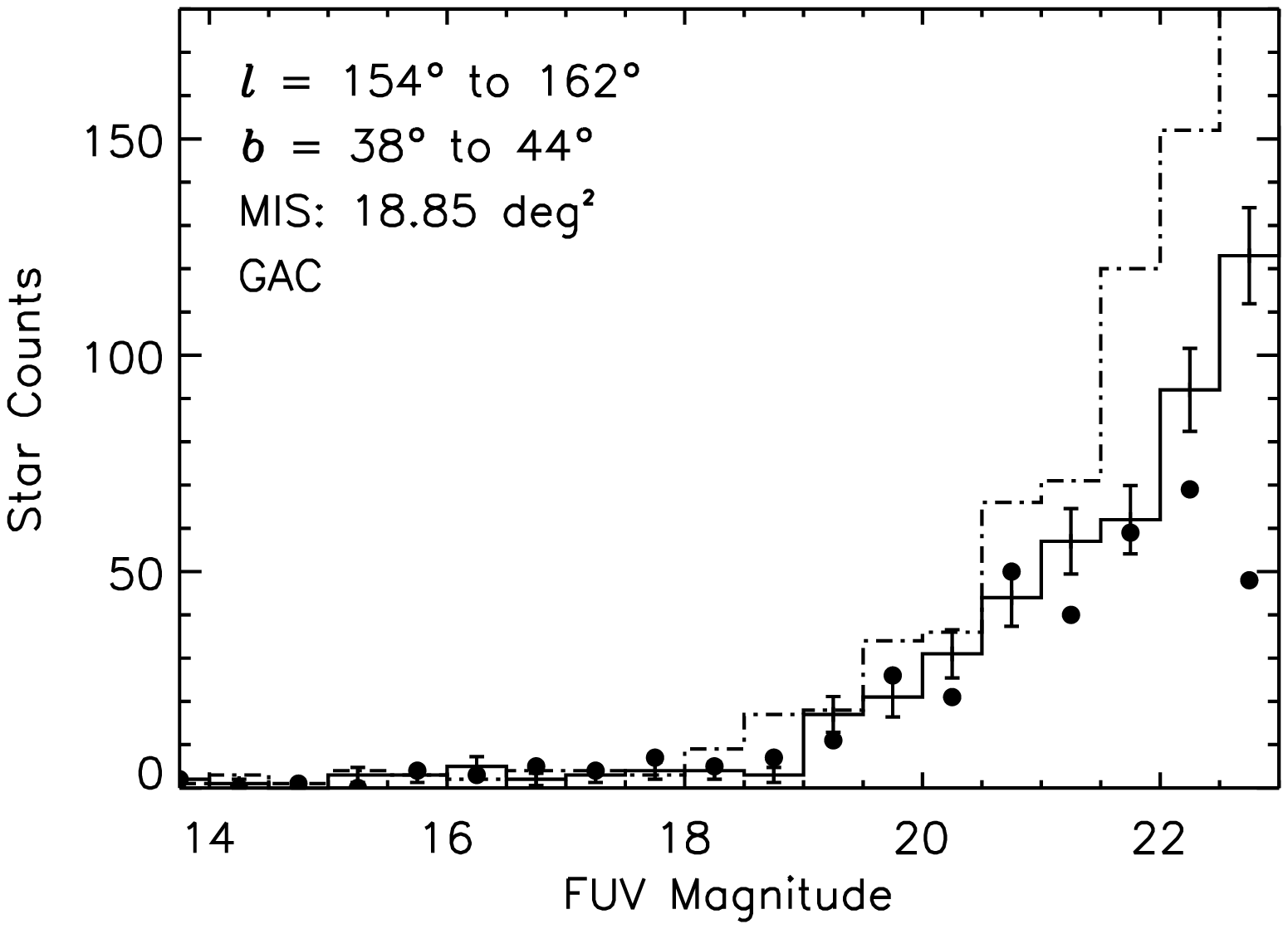}}\\

\caption{Distribution in FUV magnitudes of the UV-IR stars (solid circles) and model-predicted (solid line) star counts for the AIS and MIS fields towards the GC and the GAC (fields 5 - 8 in Table 1). The dashed-dotted line represents FUV star counts for the BaF2 band of UVIT/ASTROSAT (1350 - 1750 \AA, $\lambda_\mathrm{eff}$ = 1504 \AA). The FUV magnitudes are binned in intervals of 0.5. The model error bars shown in the plots are due to Poisson noise.
\label{fig8}
}
\end{figure*}

\begin{figure*}
\centering
 \includegraphics[width=12cm]{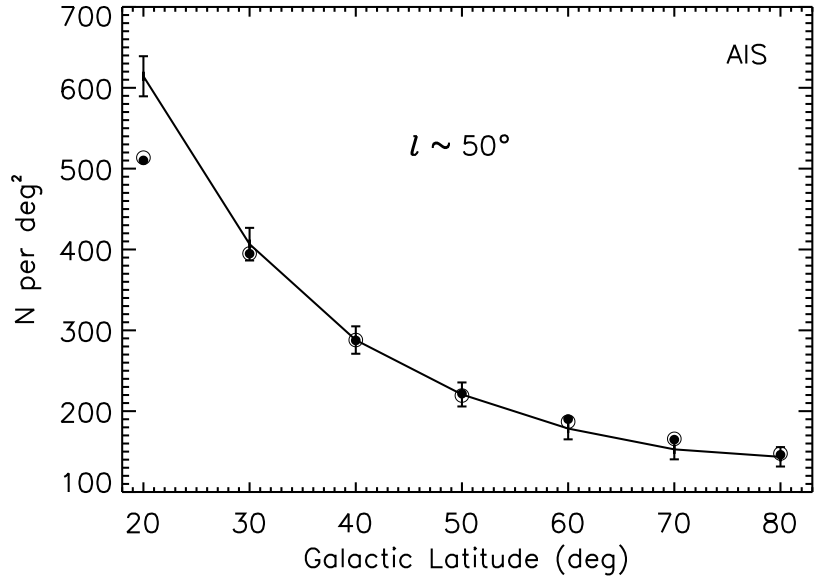}
\caption{The latitude variation of NUV star counts (per deg$^{2}$) for both the observation and model simulation at $l\sim$ 50$^{\circ}$. The UV-IR stars, GALEX+SDSS stars and model-predicted star counts are represented by solid circles, open circles and solid line, respectively. The UV-IR stars and the GALEX+SDSS stars shown in the plot are for NUV magnitude errors $<$ 0.2. The error bars displayed in the model star counts are due to Poisson noise. The asymmetric errors in the UV-IR star counts which arise due to the propagation of photometric errors are not shown.
\label{fig9}}
\end{figure*}

\begin{figure*}
\centering
\subfloat{
 \includegraphics[width=7.5cm]{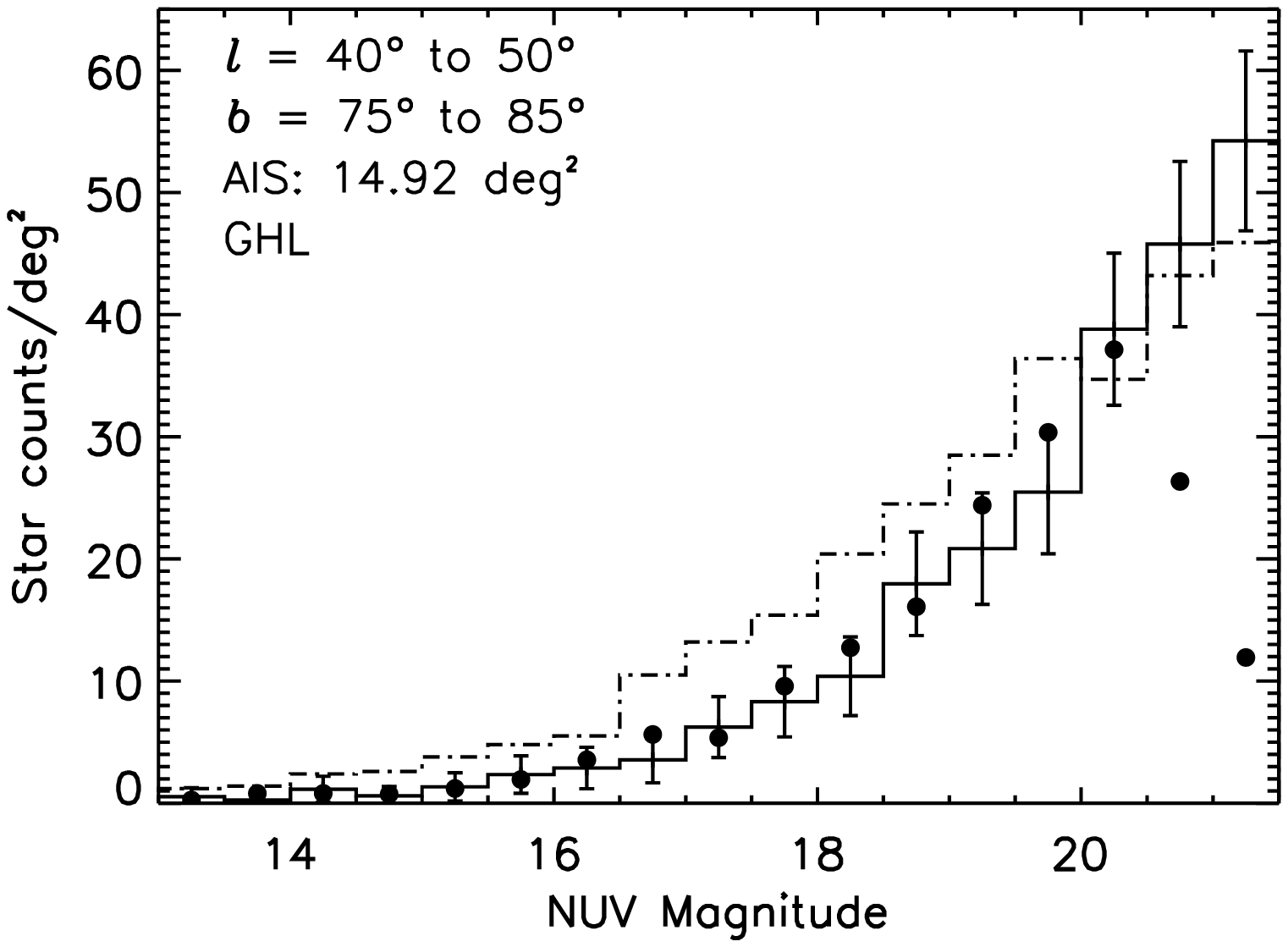}}
\subfloat{
 \includegraphics[width=7.5cm]{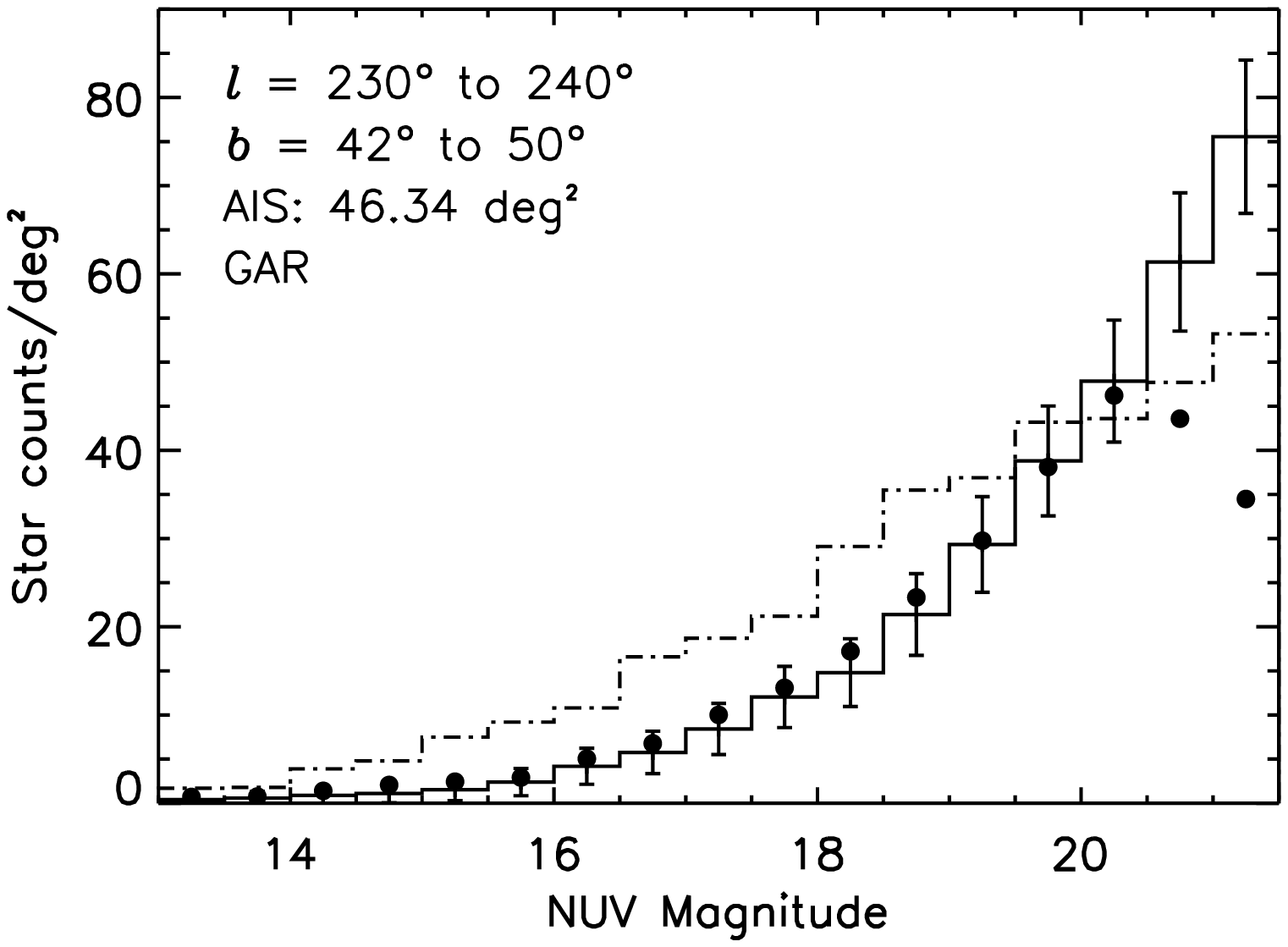}}
\caption{Comparison of NUV star counts predicted by the Besan\c{c}on (solid line) and TRILEGAL (dashed-dotted line) models of stellar population synthesis for the fields towards the GHL and the GAR. The solid circles represent the UV-IR stars. The Besan\c{c}on model error bars shown in the plots are due to Poisson noise.
\label{fig10}}
\end{figure*}

\begin{figure*}
\centering
\subfloat{
 \includegraphics[width=7.5cm]{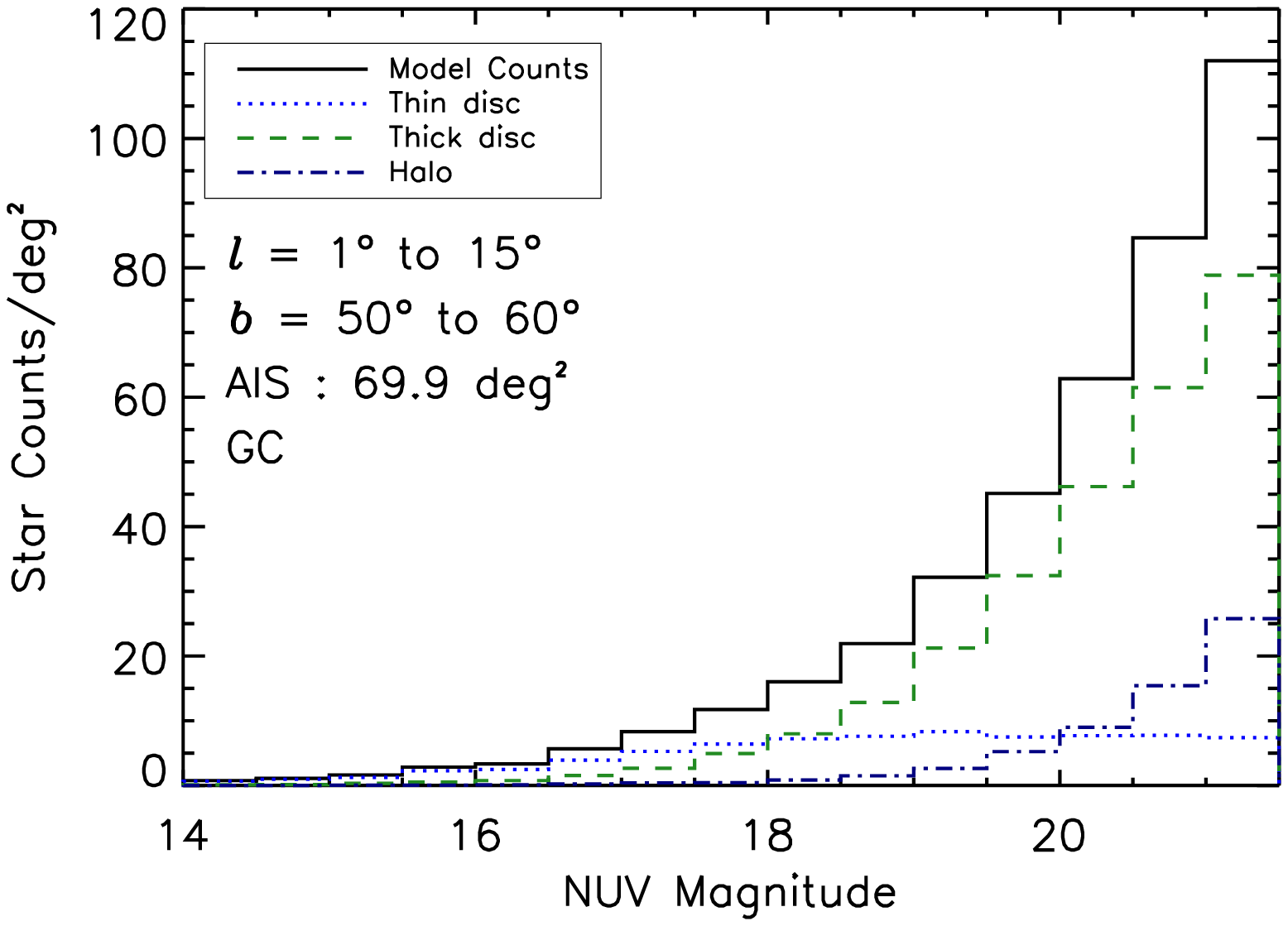}}
\subfloat{
 \includegraphics[width=7.5cm]{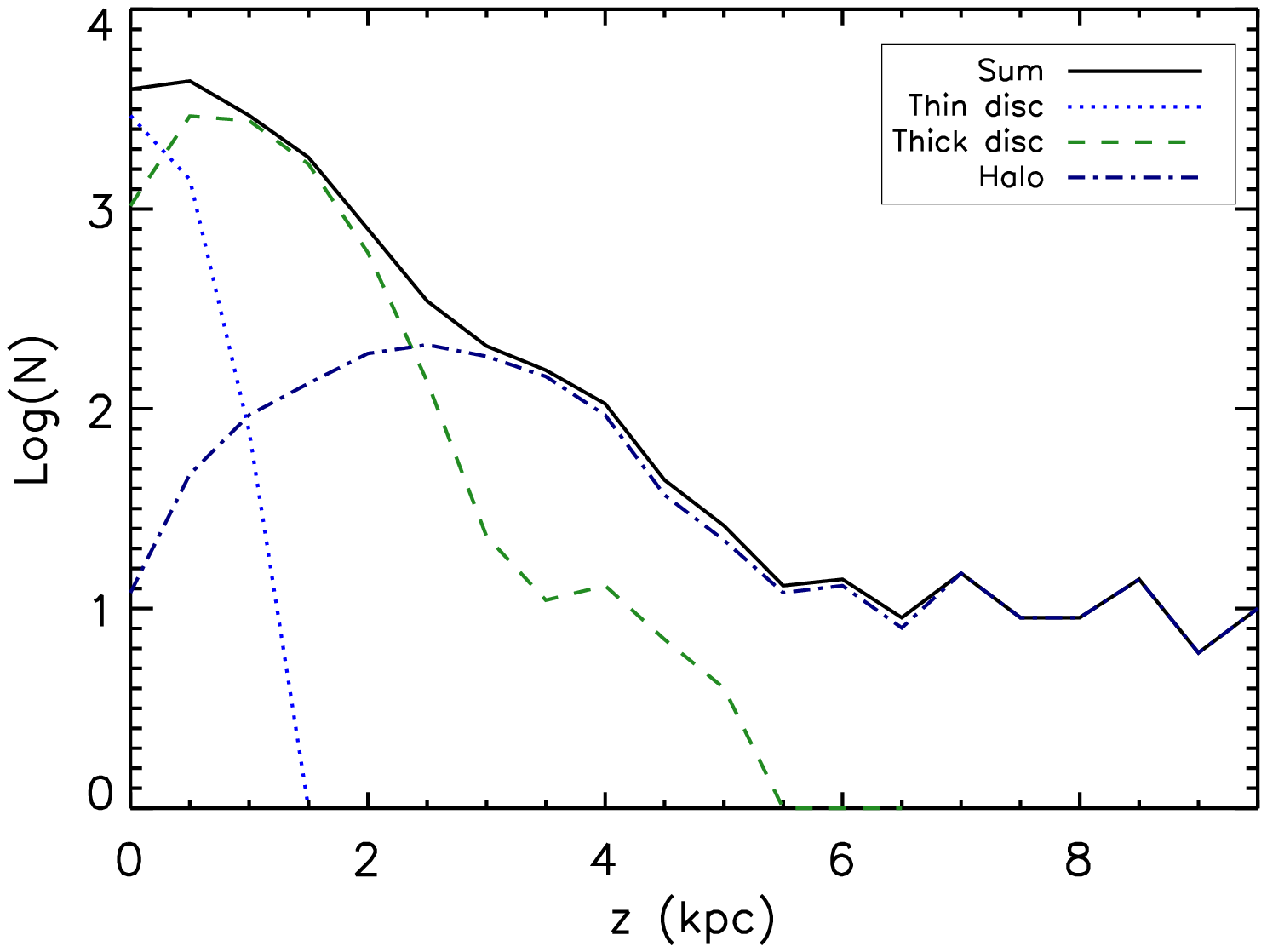}}
\caption{The left panel represents the distribution of various stellar populations produced by the Besan\c{c}on model of stellar population synthesis in the direction of GC. Different lines are explained in the legend. Similarly, the right panel represents the vertical distribution of the stellar populations in the same direction as the left panel.
\label{fig11}}
\end{figure*}
 \begin{figure*}
\centering
\subfloat{
\includegraphics[width=7.5cm]{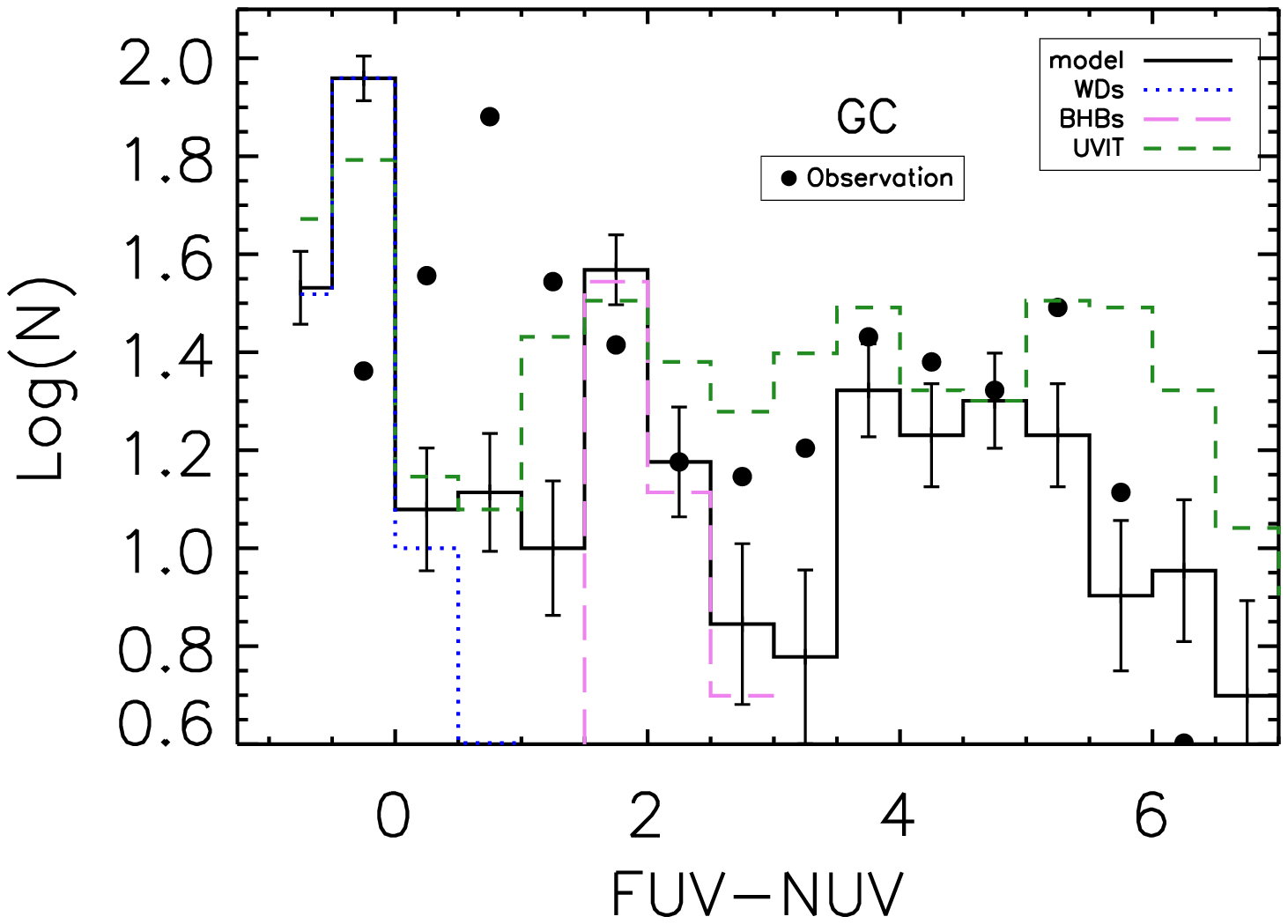}}
\subfloat{
\includegraphics[width=7.5cm]{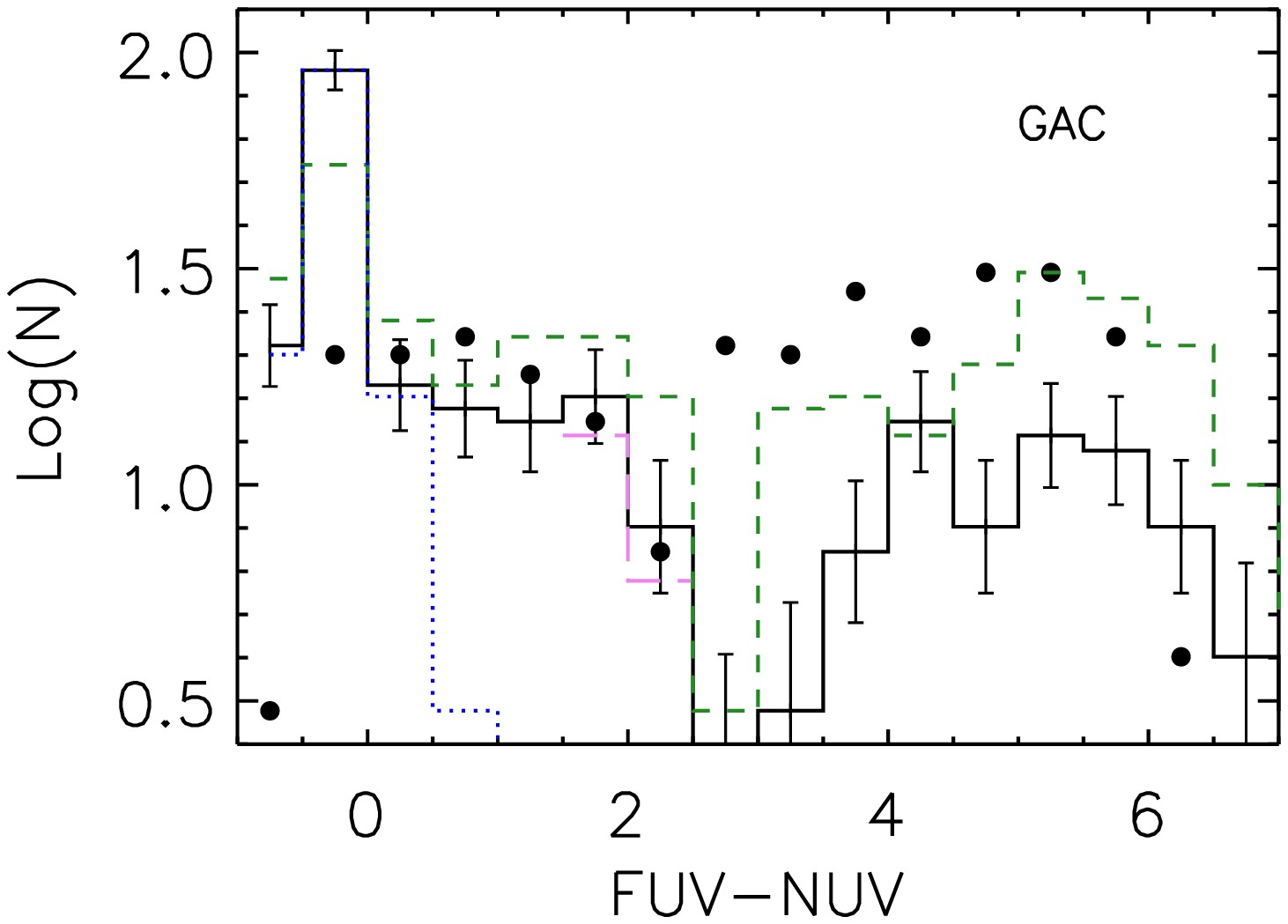}}
\caption{Comparison of $FUV - NUV$ colour between the model-predicted star counts (solid line) and the UV-IR stars (solid circles) for the AIS fields towards the GC and the GAC (fields 5 - 6 in Table 1). The UV colours shown are for the stars with FUV $<$ 20.0 magnitude, NUV $<$ 20.5 magnitude and the photometric errors $<$ 0.2 (in both FUV and NUV bands). $FUV - NUV$ colours of the model-predicted WDs and BHBs are shown by a dotted line and a long-dashed line, respectively. The UVIT ($BaF2 - NUVB4$) colour coverage is indicated by a dashed line.
\label{fig12}}
\end{figure*}
 \begin{figure*}
 \centering
\subfloat[]{
\includegraphics[width=7.5cm]{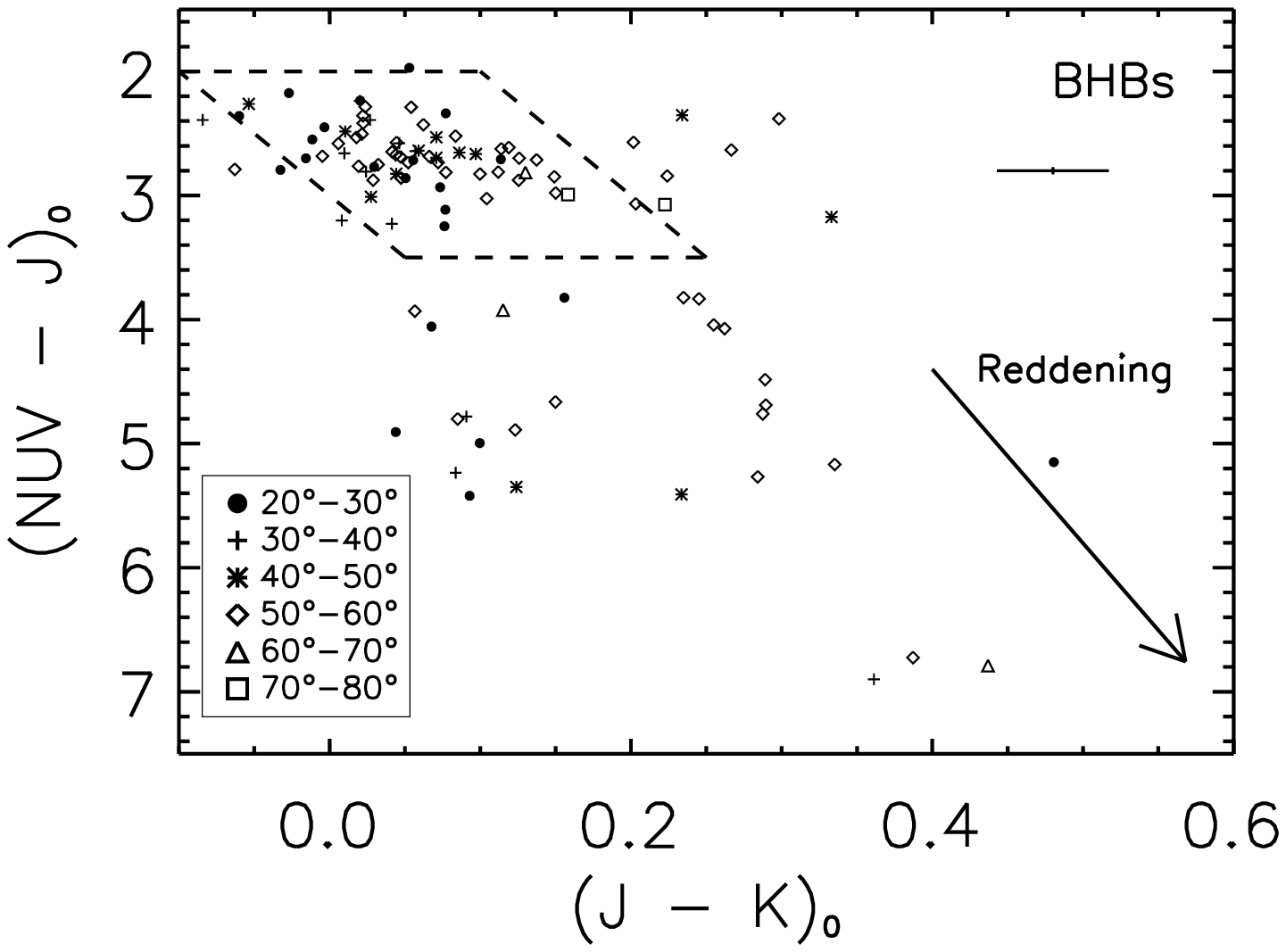}}
\subfloat[]{
 \includegraphics[width=7.5cm]{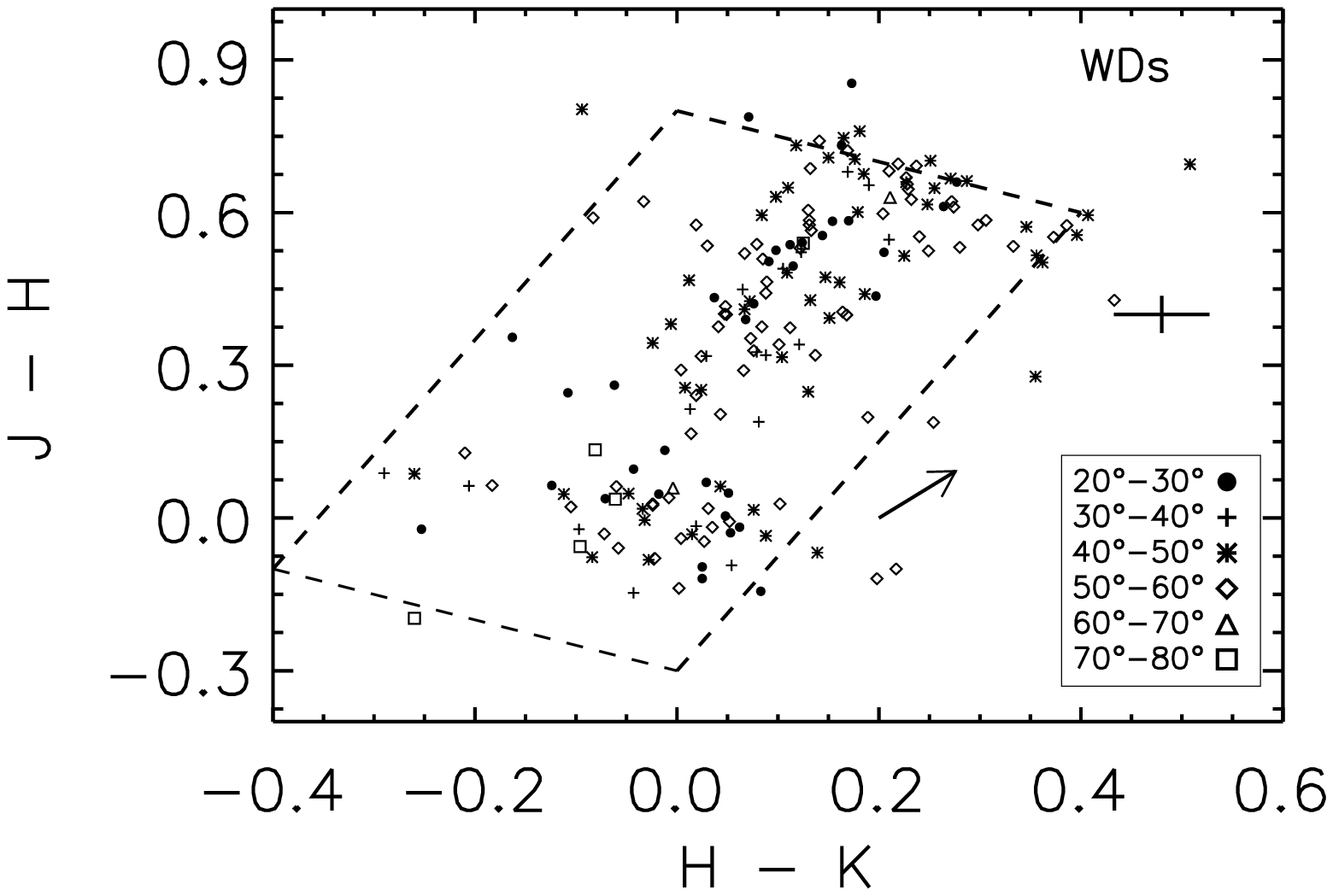}}
\caption{a) $(J - K)_{0}$ versus $(NUV - J)_{0}$ colour-colour diagram for the BHB candidates of the combined AIS fields given in Table 1. The parallelogram encloses the area occupied by the BHB samples of \citet{Kinman07}. The sources at various latitudes are represented by different symbols. b) $H - K$ versus $J - H$ colour-colour diagram for the WD candidates of the combined AIS fields in Table 1. The dashed line rectangle encloses the area used from \citet{Hoard07}. A reddening vector (the arrow) of A$_{\rm V}$ = 1 magnitude and the mean error bars of the colours are displayed in both the diagrams.
\label{fig13}}
\end{figure*}

\end{document}